\documentclass{emulateapj}

\usepackage{graphics,epsfig}

\shorttitle{Solar prominences embedded in flux ropes}
\shortauthors{Terradas et al.}

\begin{document}

\title{Solar prominences embedded in flux ropes:\\ morphological features and
dynamics from 3D
MHD simulations} 

\author{J. Terradas$^1$, R. Soler$^1$, M. Luna$^2$, R. Oliver$^1$, J.~L.,
Ballester$^1$ \& A.~N. Wright$^3$}  
\affil{$^1$Departament de F\'\i sica, Universitat de les Illes
Balears, E-07122 Palma de Mallorca, Spain and \\ Institute of Applied Computing
\& Community Code (IAC$^3$), UIB, Spain\\
$^2$Instituto de Astrofs\'\i ca de
Canarias, 38205 La Laguna, Tenerife, Spain and \\ Departamento de Astrof\'\i
sica, Universidad de la Laguna, 38205 La Laguna, Tenerife, Spain\\
$^3$School of Mathematics and Statistics, University of St Andrews, St Andrews, KY16 9SS, UK
}
\email{jaume.terradas@uib.es} 

\begin{abstract} 
The temporal evolution of a solar prominence inserted in a
three-dimensional magnetic flux rope is investigated numerically. Using the
model of \citet{titovdem1999} under the regime of weak twist, the cold and dense
prominence counteracts gravity by modifying the initially force-free magnetic
configuration. In some cases a quasi-stationary situation is achieved after the
relaxation phase, characterized by the excitation of standing vertical
oscillations. These oscillations show a strong attenuation with time produced by
the mechanism of continuum damping due to the inhomogeneous transition
between the prominence and solar corona. The characteristic period of the
vertical oscillations does not depend strongly on the twist of the flux rope.
Nonlinearity is the responsible for triggering the Kelvin-Helmholtz instability
associated to the vertical oscillations and that eventually produces horizontal
structures. Contrary to other configurations in which the longitudinal axis of
the prominence is permeated by a perpendicular magnetic field, like in unsheared
arcades, the orientation of the prominence along the flux rope axis prevents the
development of Rayleigh-Taylor instabilities and therefore the appearance of
vertical structuring along this axis.
\end{abstract}


\maketitle

\section{Introduction} 

Active region prominences are often associated to magnetic structures that seem to have
a flux rope geometry, i.e., a coherent structure in which magnetic field lines wind
around a central axis. Several attempts have been done to provide the theoretical
background, based on the equations magnetohydrostatics that describes this kind of
magnetic configurations
\citep[e.g.,][]{kuperusraadu1974,priestetal89,lowzhang04,bloklandkeppens11,hilliervan13}.
In most cases the heavy prominence is absent from the magnetic configuration
\citep[see the review of ][]{mackayetal10}, and/or the magnetic field lines do not
connect to the photosphere since the models are mostly 2D. These  models are quite
limited and restrain a proper understanding of the dynamics of suspended prominences.
Only recently, it has been possible to create a self-consistent plasma-carrying flux
rope \citep{xiaetal2014a} and to produce an in situ condensation to a prominence due to
thermal instability \citep{xiaetal2014b}.  

Nowadays it is clear that solar prominences are very dynamic, and several observed
phenomena have been identified as the result of the development of different plasma
instabilities, such as the magnetic Rayleigh-Taylor (MRT) or the Kelvin-Helmholtz (KH)
instabilities \citep[e.g.,][]{ryuetal10}. Other dynamic events affecting the whole
prominence body are the ubiquitous large amplitude oscillations associated to winking
filaments \citep[see the review of][]{tripatetal09} or the longitudinal oscillations
\citep[][]{lizhang2012,lunaetal2014}. It is evident that an accurate theoretical
description of these oscillations, usually done in terms of magnetohydrodynamic (MHD)
waves, demands realistic prominence models, and the incorporation of the 3D geometry is
essential.

Solar prominences are in general quite inhomogeneous and this property has important
consequences regarding oscillations, which are susceptible to experience resonant damping.
Here we understand by resonant absorption or resonant damping the process in which the
energy of the global oscillation is transferred to non-homogeneous layers. Since in
our problem the oscillations are not externally driven and monochromatic, the energy is not
transferred to a single position (the resonant position) in the layer. Instead, the energy
accumulates around the resonant position, i.e. where there is a perfect match between the
real part of the frequency of the global mode (often called a quasimode) and the frequency
of the local Alfv\'en modes. Importantly there is also energy transfered to frequencies
around the resonant Alfv\'en frequency. For this reason, we think that the term resonant
absorption in the case of the initial value problem is indeed not very accurate since
strictly speaking the energy is not transferred to a single resonant position. A more
general name for this mechanism of attenuation is continuum damping
\citep[e.g.,][]{sedla1995}. Another way to refer to this process is mode conversion or mode
coupling to natural Alfv\'en oscillations \citep[e.g.,][]{rae1982,leerob86,
pascoeetal10,hoodetal2013,rudermanterradas2013}. The term phase-mixing has been also used in
the literature \citep[e.g.,][]{tataronisgross1973,grossmanntataronis1973} while in
magnetospheric plasmas the term field line resonance (FLR) is usually preferred.

In prominences the strongest density inhomogeneity is usually associated to the layers that
connect the prominence core with the corona through  the prominence-corona transition region
(PCTR). Therefore, the inhomogeneous PCTR is crucial for the existence of Alfv\'en continuum
modes that are eventually responsible for the damping. Recently, the idea behind resonant
damping has been applied to prominences/threads 
\citep{arreguiterretal08,soleretal09c,arrball10,soleretal10,arreguietal11,okamotoetal2015,antolinetal2015}.
The continuum damping is not new, and has been investigated in the past under different
frameworks; laboratory plasmas
\citep[][]{tataronisgross1973,chenhasegawa1974,poedtsetal1992}, coronal loop oscillations
\citep[e.g.,][]{ionson78,hollweg87,hollyang88,sakuraietal91,goossensetal92,goossetal02,
rudrob02}, and magnetospheric plasmas \citep[][]{southwood1974,mannetal1995,wrightrick1995}.
In the previous studies both the driven and the initial value problem have been addressed,
and the term resonant absorption has been used in most of the cases except for
magnetospheric plasmas. For historical reasons we tend to keep using the name resonant
absorption. 

With the goal of improving the existing magnetohydrostatic prominence models, \citet{terradasetal2015}
have studied the morphology of a prominence suspended in a 3D arcade configuration. In that work, the
authors have found that the prominence is specially prone to develop MRT instabilities \citep[see
also][]{hillieretal2012a,hillieretal2012b}. In the present study we extend the work of these authors to
a twisted flux rope, more representative of active region prominences rather than quiescent prominences.
In particular, we choose the magnetic configuration constructed by \citet{titovdem1999}. This
three-dimensional magnetic model can represent a wide variety of flux ropes depending on the parameters,
and has the advantage that it is easily implemented using analytic expressions for the 3D magnetic field.
The aim of this paper is to understand the main morphological features of a prominence embedded in a
flux rope for different values of the magnetic twist. The models studied here are global (the fine
structures of the prominence, i.e., threads, are not resolved) and could be used, in future studies, to
analyze large amplitude oscillations associated to winking filaments or to longitudinal oscillations.  

This study is challenging from several points of view. Firstly, the process of resonant
absorption, associated to the oscillations, has the particularity that decreasing spatial
scales are continuously built with time through phase-mixing \citep[e.g.,][]{heypri83}. This
inevitably leads to a situation in which the grid resolution is unable to capture the small
spatial scales and this produces an artificial (numerical) dissipation of energy in the
system. We have conceived specific numerical experiments to understand the effect of
numerical dissipation on the results. The process of wave leakage, i.e., the emission of
fast magneto-acoustic waves from the prominence, can also operate in our model and produce a
physical energy loss from the system. We have also devised particular simulations, based on
the application of special boundary conditions, to investigate this problem. Secondly, and
closely related to the first point, is the presence of instabilities. In particular, for the
configuration studied here KH instabilities due to the strong shear velocities at the
prominence edges play a relevant role in the morphology of the structure. Eddies are
generated at the PCTR and energy cascades to small spatial scales until numerical
dissipation becomes important. Another complication in this study is that the numerical
techniques applied to capture the vigorous motions associated with the KHI may impede a
proper analysis of possible stationary states of the system. This issue is properly treated
in this work by the use of specific numerical techniques. Thus, the problem investigated
here involves several ideal mechanisms that are not independent one from another and
generate a rich dynamism in the system, but at the same time we intend to find out potential
stationary regimes.

\section{Initial setup}\label{init}

\begin{figure}[!ht] \center{\includegraphics[width=5.cm]{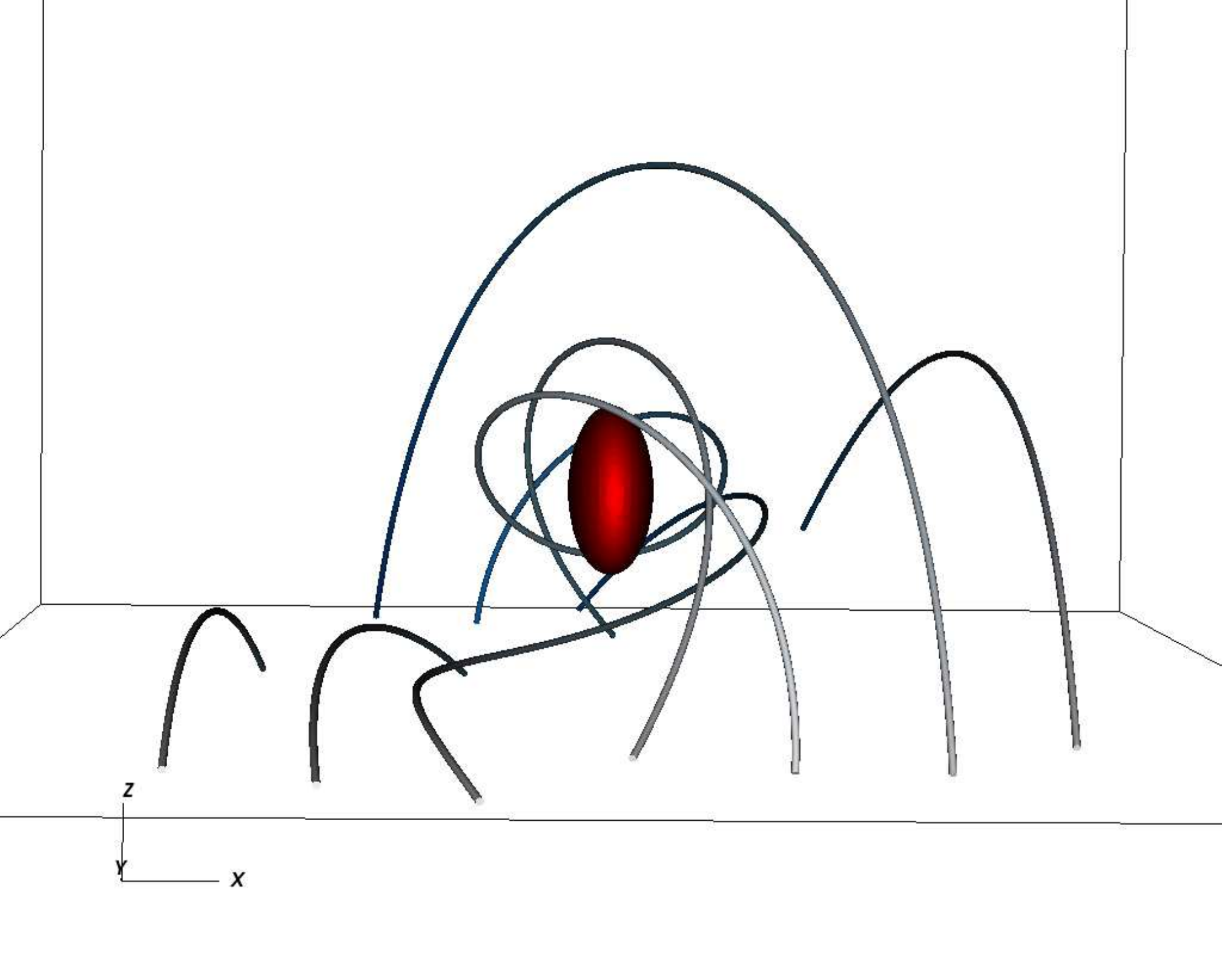}}
\center{\includegraphics[width=5.cm]{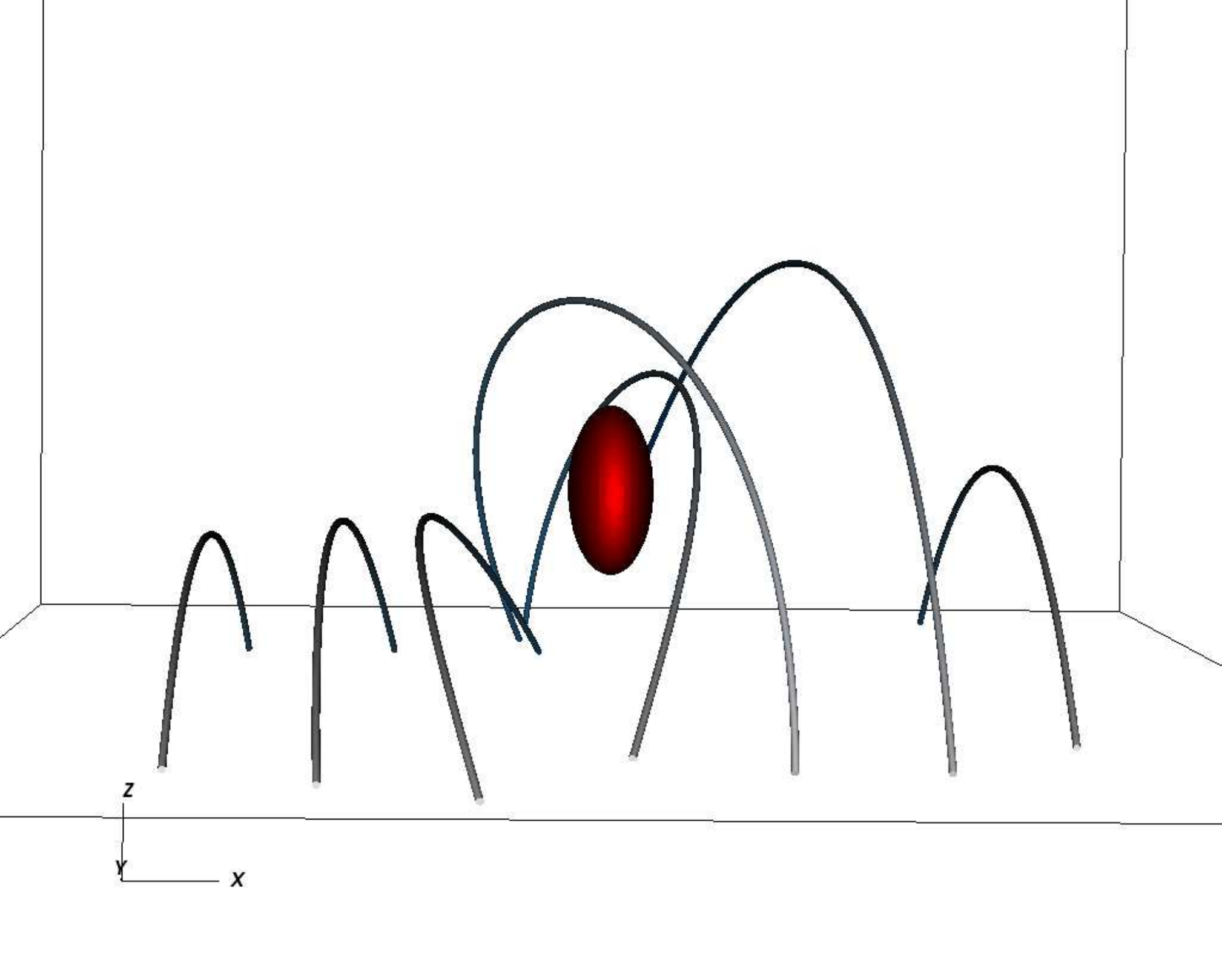}}
\center{\includegraphics[width=5.cm]{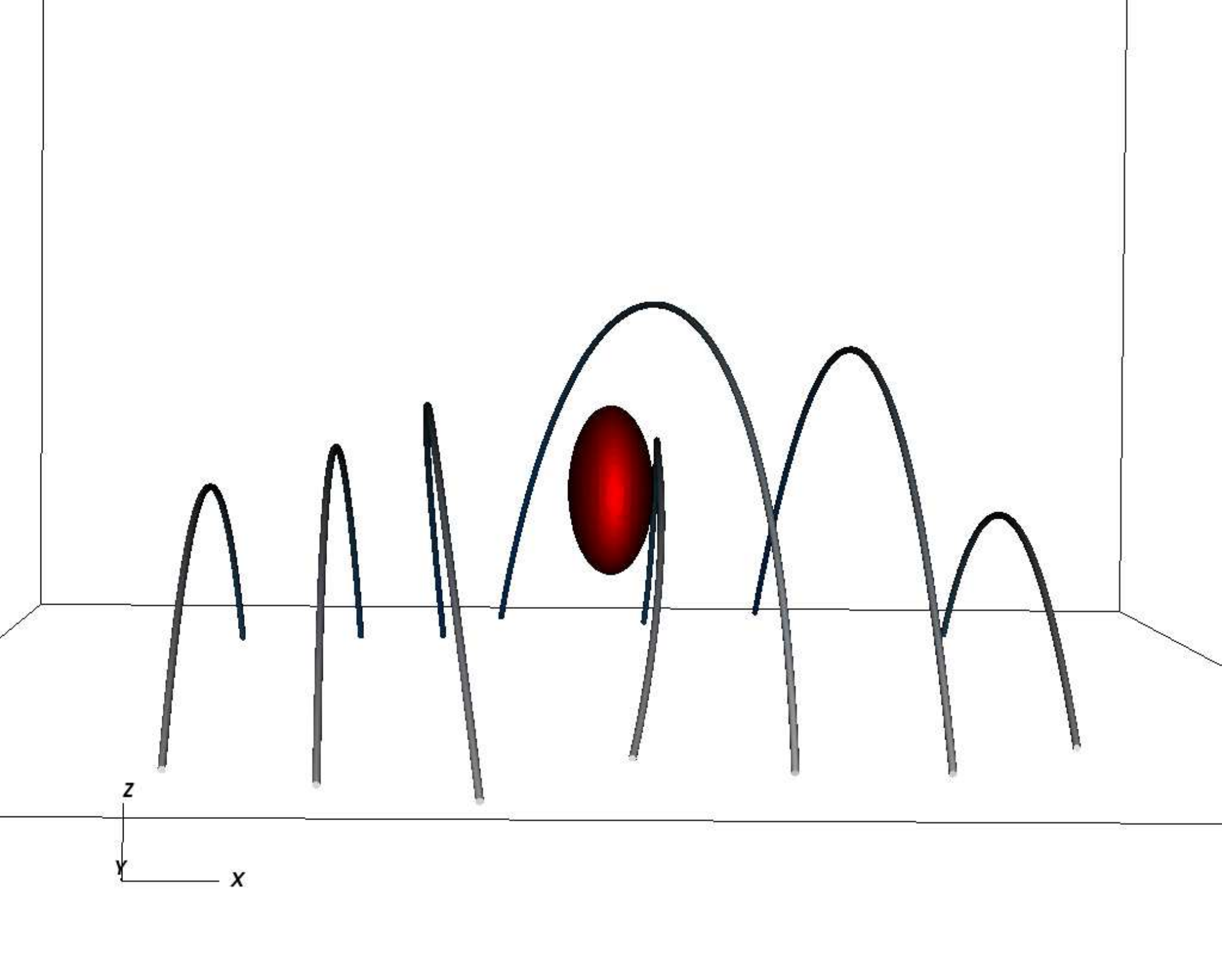}}\caption{\small Magnetic
field configuration and initial density enhancement (in red colors) for
$N_t=8.5$ (top panel), $N_t=4.3$ (middle panel), and $N_t=2.2$ (bottom panel).
The same magnetic field lines rooted at the base of the corona have been
selected in the three plots.}\label{magfield} \end{figure}

In our model the background configuration is a vertically stratified atmosphere due to
constant gravity at a temperature of $1\,\rm MK$, representative of coronal conditions. This
atmosphere is permeated by a magnetic field following the \citet{titovdem1999} model, which
is a three-dimensional toroidal force-free model that contains a poloidal component of the
magnetic field. In the construction of the model it is assumed that the small radius of the
flux rope is much smaller than the large radius of the toroid. This model has been used in
the past mainly to study eruptions and coronal mass ejections (CME)
\citep[see][]{torokkliem2003,toroketal2004}. Here we concentrate on a stable configuration
with respect to the kink instability that appears when the twist is above a given threshold.
The amount of twist is controlled by the parameter $N_t$. Values of 8.5 (strong), 4.3
(moderate) and 2.2 (weak twist) are considered in this work. It is worth mentioning that for
the strongest twist, the ratio between the toroidal and azimuthal magnetic field component
is around 2 at the edge of the flux rope but the field lines do not have more than three
turns, in agreement with observations of stable flux ropes. Examples of the three magnetic
configurations are found in Figure~\ref{magfield}. Note that for  $N_t=4.3$ and 2.2 the
configurations resemble arcades whose magnetic shear increases with height. In the three
models the small radius of the toroidal flux rope is $a=10\,\rm Mm$, while the big radius is
$R=140\,\rm Mm$. The height of the center of the flux rope is $20\,\rm Mm$.

The prominence is represented by a three-dimensional density enhancement with respect
to the background density, and is located inside the flux rope. A simple Gaussian
profile in each direction is used. The characteristic dimensions of the prominence are:
width of $5\,\rm Mm$, height of $10\,\rm Mm$, and a length of $20\,\rm Mm$. The center
of the prominence is initially situated at $15\,\rm Mm$ from the base of the corona.
This density enhancement is also represented in Figure~\ref{magfield} in a 3D view
pointing along the longitudinal axis of the prominence, i.e., along the $y-$direction.
Different values for the density enhancement, and therefore total masses of the
prominence, have been considered allowing us to study situations of extremely light
prominences to cases with total masses similar to the ones inferred from observations.
In particular, the maximum density enhancement used in this work is 30 times the
coronal density. Although this value is a bit low in comparison with the densities
estimated  from observations, it gives a total mass of $1.8\times 10^{10}\,\rm kg$,
which is typical of light prominences. Larger density enhancements, and therefore total
masses, can lead to a situation in which the magnetic structure is unable to sustain
the prominence \citep[see][]{terradasetal2015}. Increasing the magnetic field provides
additional magnetic support to the prominence since the magnetic force scales as $B^2$.
For example, if we double the magnetic field the Lorentz force is four times larger and
therefore a prominence four times more massive could be sustained by this new field.
The intensity of the magnetic field determines, together with the density, the value of
the Alfv\'en speed. In the solar corona this value is at most of the order of
$1.8\times 10^{3}\,\rm km\, s^{-1}$, meaning that the magnetic field is constrained by
this velocity. In our simulations  the maximum value of $B_z$ at the base of the corona
is 5 Gauss, and it fulfills the restriction given by the typical coronal Alfv\'en speed.
With this magnetic field and the corresponding gas pressure of a plasma at $1\,\rm MK$,
the plasma-$\beta$ is in the range $0.02-0.09$. We are clearly in a low-$\beta$ regime
and this guides the dynamic processes that take place in the configuration. Stronger
magnetic fields, which are feasible because we are considering an active region
prominence, would augment too much the value of the Alfv\'en speed and decrease the
already low value of the plasma-$\beta$.

Regarding the geometry of the magnetic field it is worth mentioning that the magnetic
support of the prominence when the longitudinal axis is essentially along the magnetic
field, such as in a flux rope with low twist like the one in the bottom panel of
Figure~\ref{magfield}, is less efficient than a situation in which the magnetic field is
perpendicular to this axis. This would correspond, in our configuration, to a
prominence with the longitudinal axis pointing in the $x-$direction. The reason is that
in this case the prominence is permeated by more magnetic field meaning that the total
magnetic force is stronger. This simple geometrical feature of our model has important
consequences with respect to the total amount of mass that can be sustained by our
configuration and, as we show later, regarding the MRT instability. In situations with
stronger magnetic twist, as in top panel of Figure~\ref{magfield}, the magnetic support
increases due to the presence of local magnetic dips but since we do not consider more
than three turns of field lines this additional support is not significantly
enhanced in our configuration.

\section{Method and boundary conditions}\label{method}

The background and the magnetic field are initially in equilibrium. Given an
initial localized density distribution representing the prominence, which is not
in equilibrium with the environment, the system is allowed to evolve with time.
To investigate the dynamics of our configuration the ideal three-dimensional MHD
equations are numerically solved. The equations are the following, 
\begin{eqnarray*}
\frac{\partial{\rho}}{\partial t}+\nabla\cdot \left({\rho \bf v}\right) =0,
\end{eqnarray*} \begin{eqnarray*} \frac{\partial{\rho \bf v}}{\partial
t}+\nabla\cdot \left({\rho \bf v \bf v} +p {\bf I}-\frac{\bf B \bf
B}{\mu}+\frac{{\bf B}^2}{2 \mu} \right) =\bf \rho g, \end{eqnarray*}
\begin{eqnarray*} \frac{\partial{\bf B}}{\partial t}=\nabla \times \left(\bf v
\times \bf B \right),   \end{eqnarray*} \begin{eqnarray*}
\frac{\partial{p}}{\partial t}+\nabla\cdot \left({\gamma p \bf v}\right) =
\left(\gamma-1\right) {\bf v}\cdot \nabla p, \end{eqnarray*}

\noindent where $\bf I$ is the unit tensor, $\bf g$ is the gravitational
acceleration, and the rest of the symbols have their usual meaning. The
nonlinear MHD equations are solved in a Cartesian coordinate system. The gravity
force points in the negative $z-$direction and the longitudinal axis of the
prominence is along the $y-$direction (see Figure~\ref{magfield}). Our model does
not include a density transition between the corona and the photosphere and the
plane at $z=0$ represents the base of the corona. The dimensions of the domain
are $-50\,{\rm Mm} <x< 50\,{\rm Mm}$, $-100\,{\rm Mm} <y< 100\,{\rm Mm}$, and
$0<z< 60\,{\rm Mm}$. Different numerical resolutions have been considered, and
the best resolution achieved with the present computational resources is
$300\,\rm km$.

The previous equations are solved using the code MoLMHD that has been updated to  be
efficient in achieving stationary solutions but at the same time being robust in the
treatment of weak shocks. With this goal in mind, the order of the spatial
derivatives has been raised to six, and the scheme has been combined with a 5th order
WENO method. The rest of the details about the code can be found in
\citet{terradasetal2015} and references therein.

Line-tying boundary conditions are applied at the base of the corona, meaning
that the three components of the velocity, {\bf v}, are set to zero, the
magnetic component perpendicular to the boundary is kept constant, $B_z$ in our
case, and the rest of the variables have their spatial derivatives equal to
zero. For the rest of the boundaries in the computational box, i.e., lateral and
top planes, two types of conditions are investigated; perfectly reflecting
boundaries, i.e., line-tying  (this condition is referred as closed boundary
conditions), and non-reflecting boundaries. This last condition is  implemented
through the method of the perfectly matched layer (PML). The method is based on
the work of \citet{berenger1994,hu2001} \citep[see also][]{parchkoso2007}, and
it is efficient in absorbing outgoing waves. Hereafter, these boundary
conditions are referred to as open boundary conditions.

\begin{figure}[!ht] \center{\includegraphics[width=7.cm]{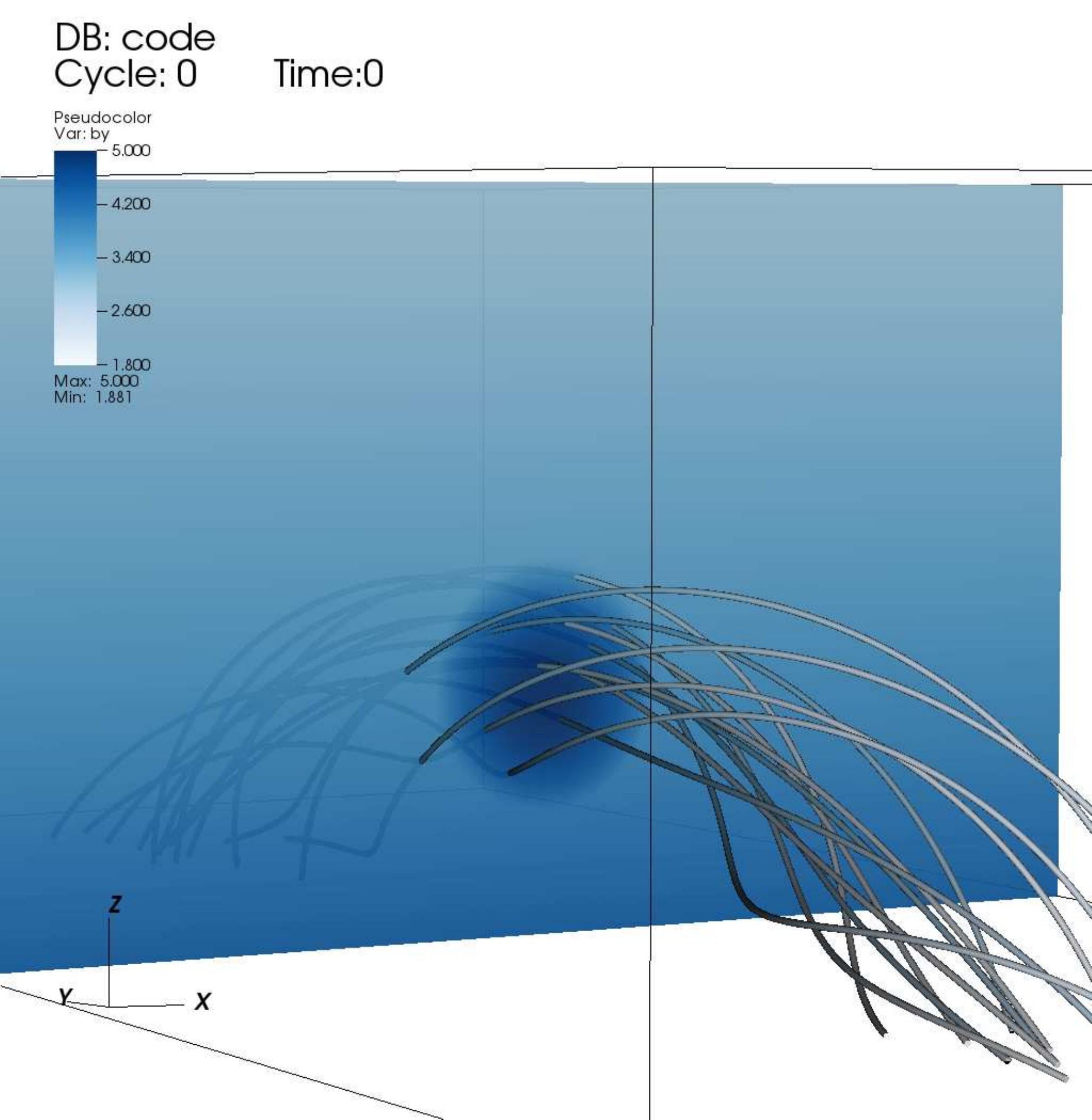}}
\center{\includegraphics[width=7.cm]{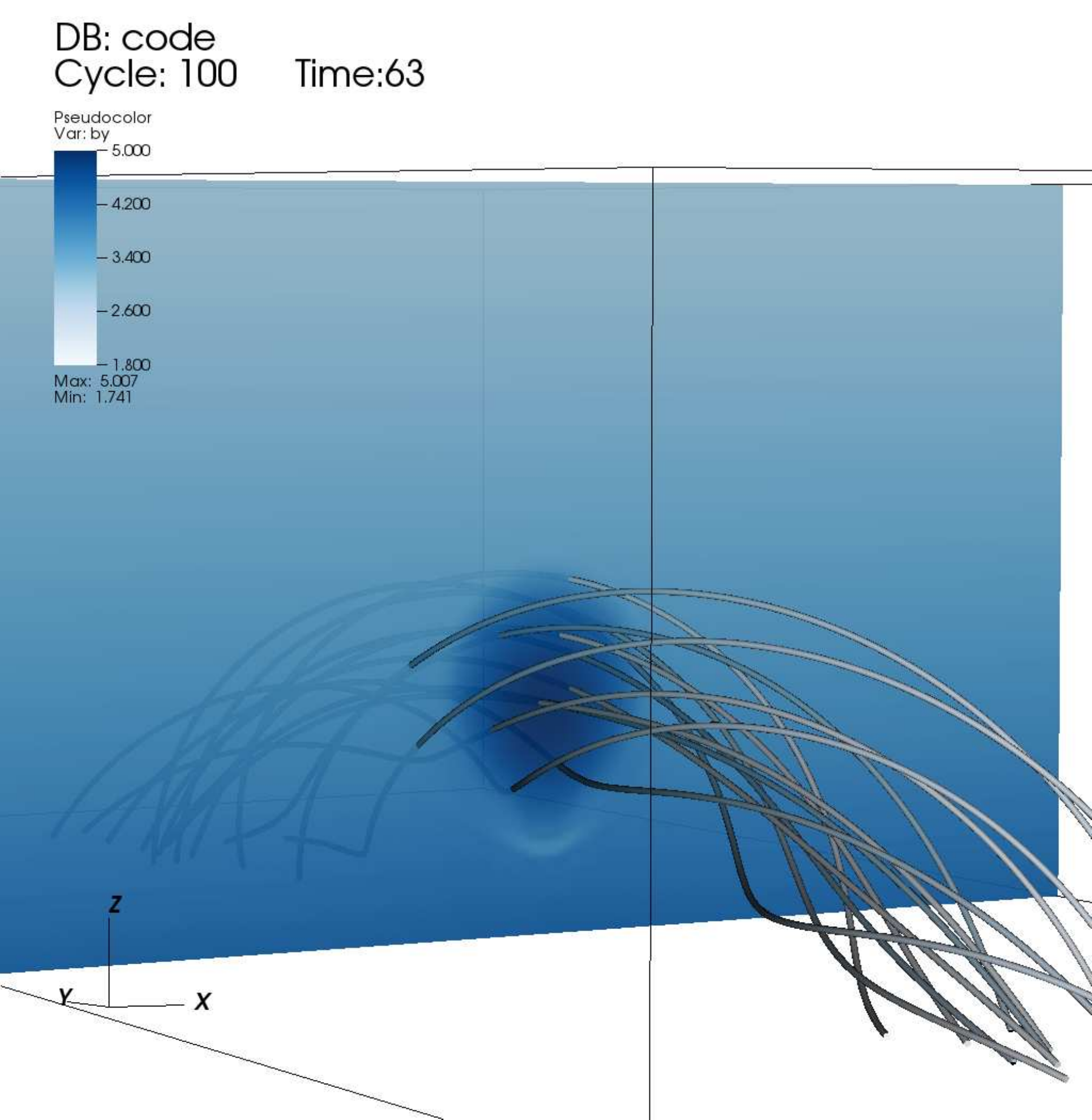}} \caption{\small
Snapshot of the flux rope configuration for $N_t=8.5$ (strong twist). The $B_y$
component of the magnetic field is plotted a the central plane together with
some specific magnetic field lines. The upper panel corresponds to the test case
without the prominence enhancement, while in the lower panel the dense
prominence inside the flux rope is present (but not visible in this
plot).}\label{noprom} \end{figure}

\section{Test case}\label{testcase}

We start with the configuration explained in Section~\ref{init} but we do not
introduce at this stage a prominence inside the flux rope. This allows us to
verify that we have correctly implemented the force-free Titov and Demoulin
model since the structure should be close to a static equilibrium. The flux rope
with the strongest twist ($N_t=8.5$), and therefore strongest current is chosen.
Closed boundaries, i.e., line-tying conditions at all the sides of the
computational box, are imposed.

In Figure~\ref{noprom}a some magnetic field lines are represented together with
the toroidal component of the magnetic field ($B_y$) at the central plane of the
configuration. The results of the evolution, not shown here, indicate that the
system is in essentially the same state as in the beginning of the simulation
meaning that the magnetic field model is indeed force-free. We find rather small
flows, of the order of $2\,\rm km\, s^{-1}$, associated to the relaxation process
of the magnetic configuration, that are eventually attenuated. The flux rope
cross section keeps the initial circular shape, indicating that the thin tube
approximation used to construct the toroidal model works well ($a/R=0.071$ in
our model). Evidences of kink or helical instabilities, that eventually would
destroy the equilibrium, are not present in the simulation. This is the expected
behavior since we have intentionally imposed a small number of turns of the
magnetic field lines around the flux rope axis.

The presence of the flux rope in the global arcade configuration produces an
increase in the local Alfv\'en speed ($v_A=B/\sqrt{\mu_0 \rho}$). It is known
that to have an efficient waveguide that is able to trap energy, a minimum in
the Alfv\'en speed is required. This means that without a dense prominence
inside the flux rope, which will produce an important decrease of the Alfv\'en
speed, the toroidal magnetic structure with just the background density is
unable to act as a wave guide, at least for fast MHD transverse waves.

\section{Flux rope with an embedded prominence}

\subsection{Global features}

The situation is different when a heavy prominence is incorporated to the flux
rope. Figure~\ref{noprom}b shows the results after one hour of evolution, and
several differences with respect to the simulation described in
Section~\ref{testcase} are evident (compare with Figure~\ref{noprom}a). The
cross-section of the flux rope is no longer purely circular and shows an oval
shape. The prominence is located initially at $z_0=15\,\rm Mm$ and because of the
gravity force the magnetic structure together with the prominence are pushed
downwards. The results of this simulation show that the upper part of the
flux rope is essentially unaltered while the lower part, where the prominence
was introduced, suffers the strongest change in the toroidal magnetic field. A
close inspection of  Figure~\ref{noprom}b reveals that $B_y$ underneath the
prominence shows a significant decrease. Note also that from  Figure~\ref{noprom}b
the magnetic field lines inside the flux rope are slightly different with
respect to Figure~\ref{noprom}a because of the presence of the prominence.

\begin{figure}[!ht] \center{\includegraphics[width=8.cm]{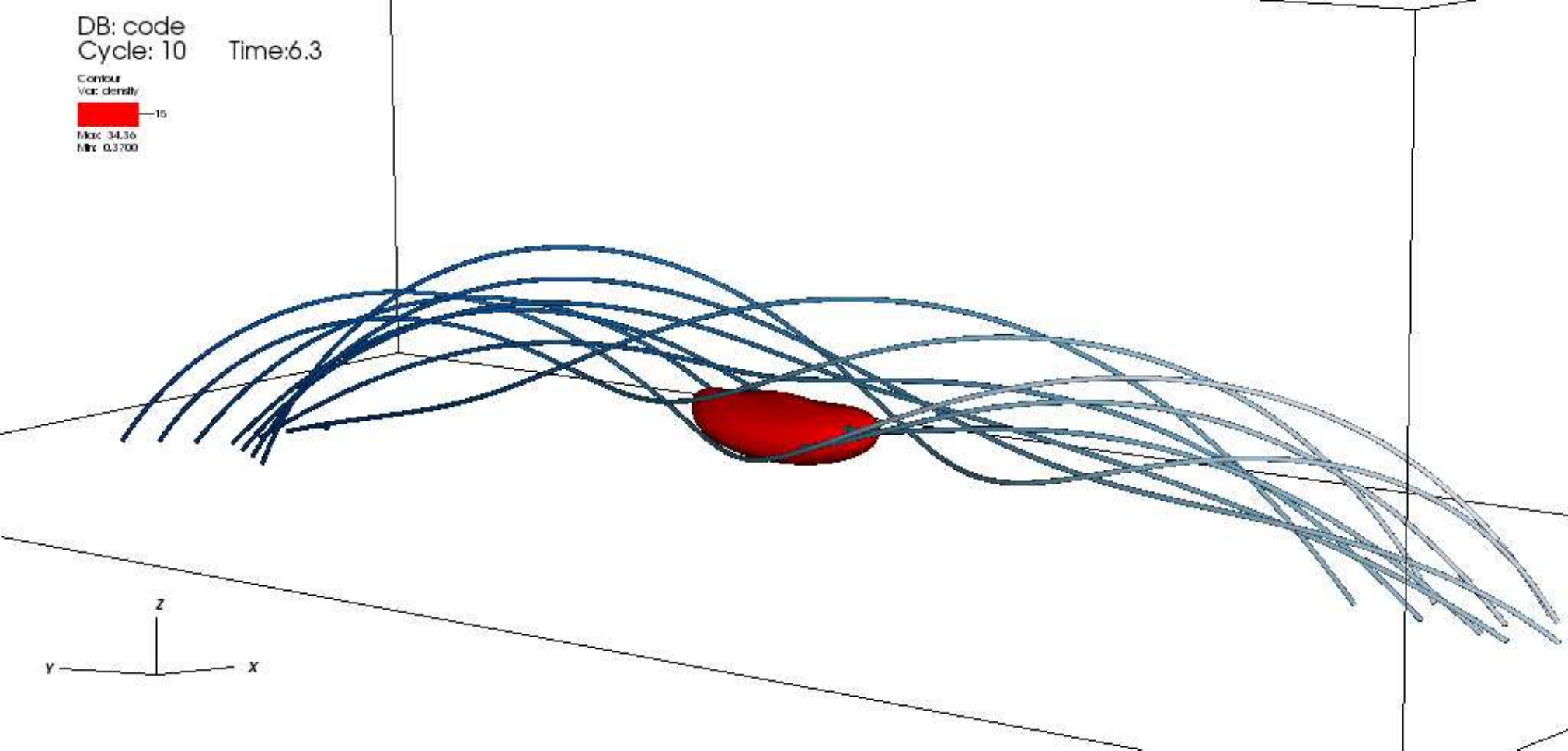}}
\center{\includegraphics[width=8.cm]{{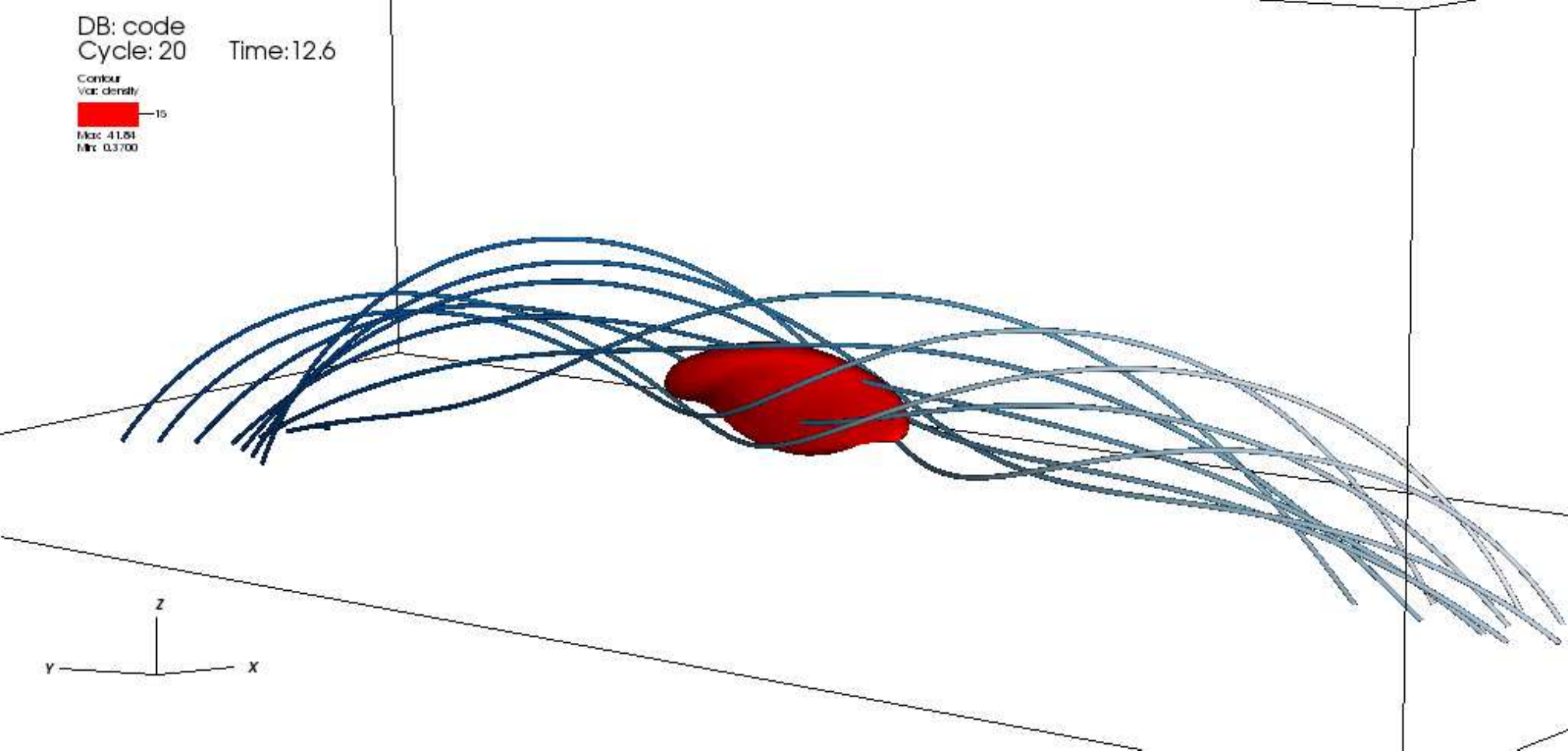}}
\center{\includegraphics[width=8.cm]{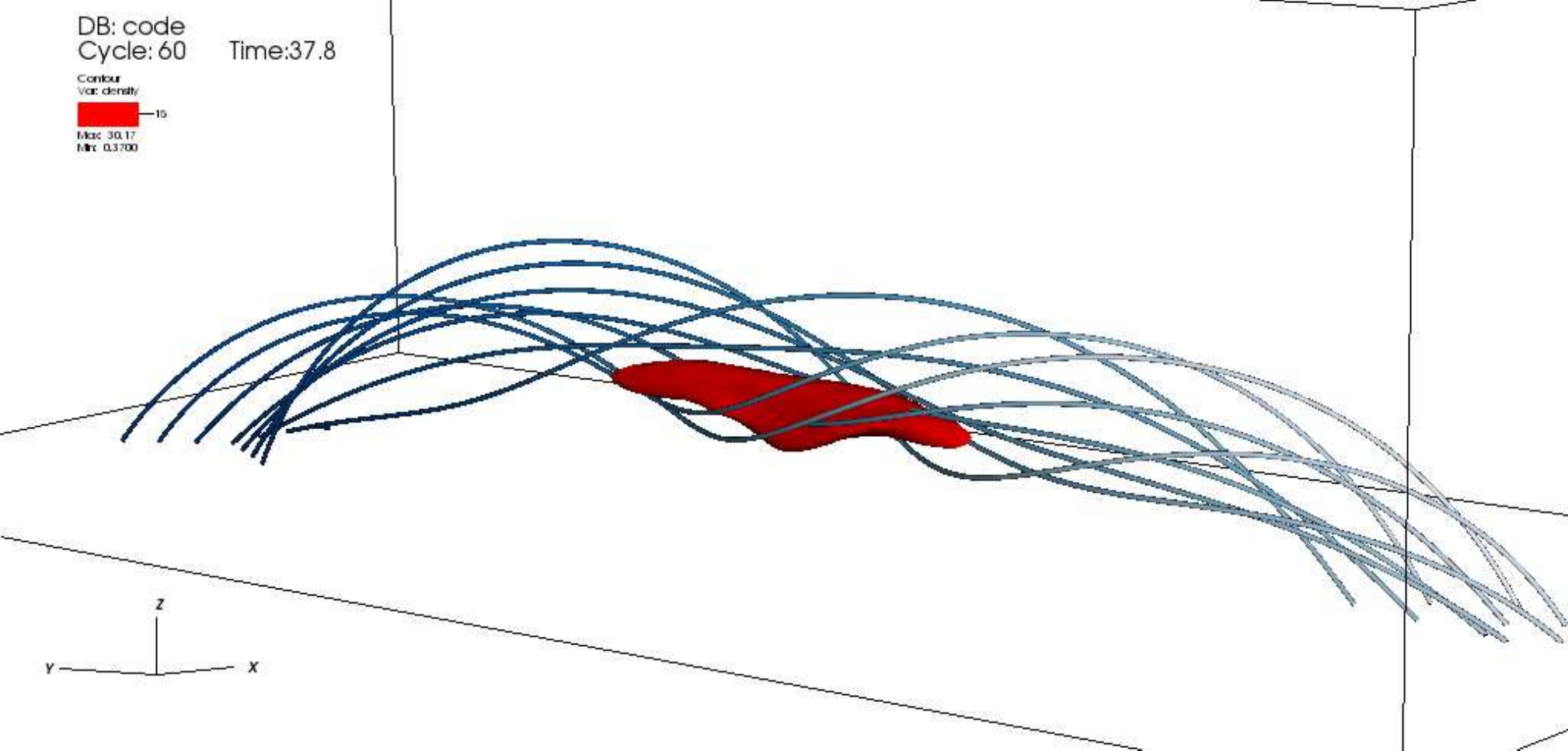}} } \caption{\small
Snapshots of the flux rope configuration plus the prominence at three different
times. For this case $N_t=8.5$, and the spatial resolution is $600\,\rm km$.  See
movie in the online version of the journal.
}\label{prom0} \end{figure}

\begin{figure}[!ht] \center{\includegraphics[width=8.cm]{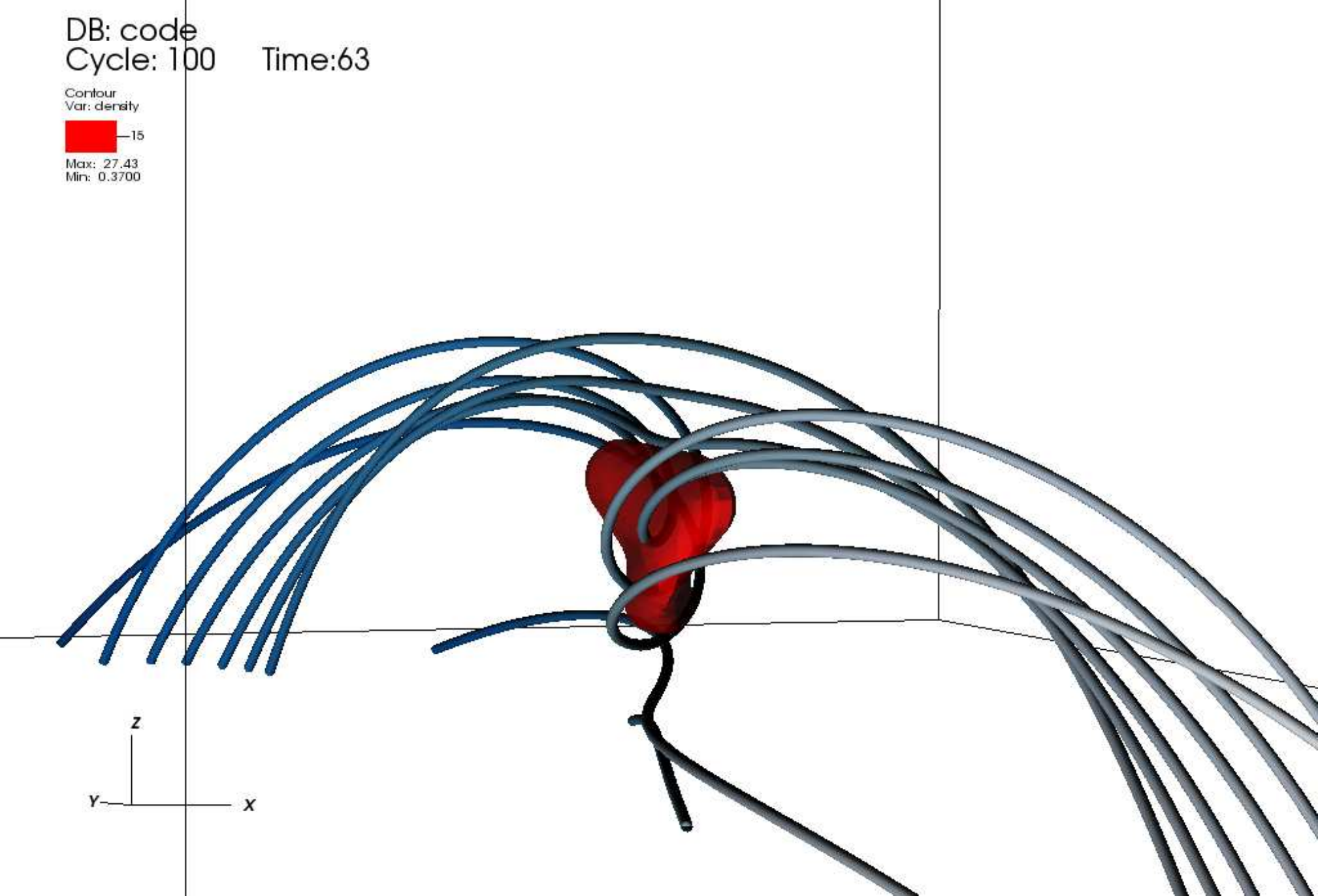}}
\center{\includegraphics[width=8.cm]{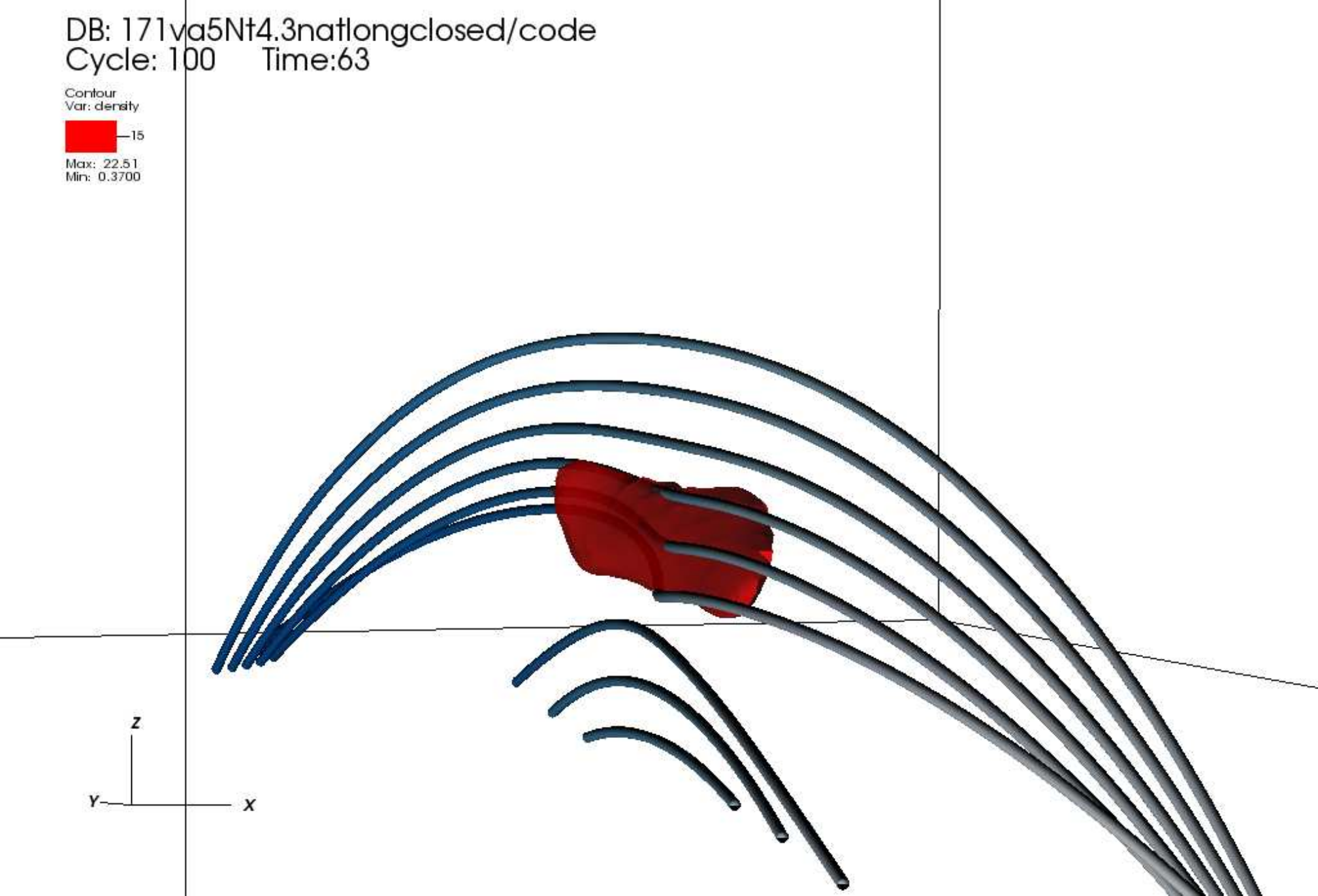}} 
\center{\includegraphics[width=8.cm]{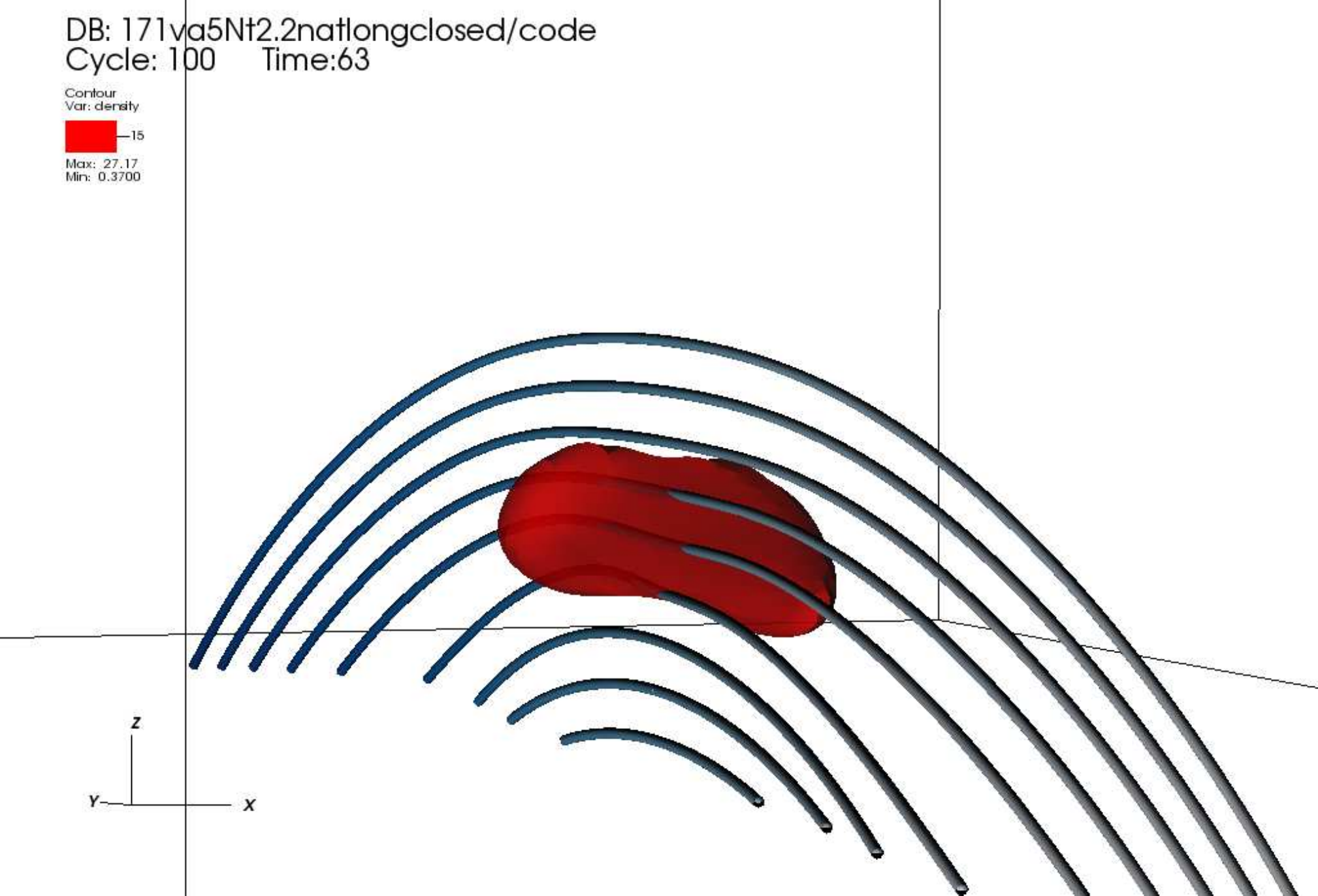}}  \caption{\small Snapshots of
the flux rope configuration plus the prominence at the same instant for the cases
$N_t=8.5$ (top panel), 4.3 (middle panel), and 2.2 (bottom panel). The KH instability
is absent here because the spatial resolution is not high enough ($600\,\rm km$).}\label{prom2} \end{figure}

A density isocontour and some magnetic field lines are represented in
Figure~\ref{prom0} at three different times. The shape of the prominence changes
from the initial Gaussian profile to a structure which is more irregular and
elongated. In this simulation the prominence is suspended above the base of the
corona mainly due to the magnetic restoring forces associated to the dips
introduced by twist that counteract gravity. The mass tends to be
aligned with the flux rope axis.

In our model the temperature at the core of the prominence is typically around 
$2.6\times 10^{4}\,\rm K$ and connects smoothly with the much hotter coronal
environment, at $1\times 10^{6}\,\rm K$, through the PCTR. The size of the transition
is mostly determined by the initial shape of the density enhancement. If the
initial prominence density is reduced then the temperature at the core of the
prominence is increased. This effect is easily understood by assuming that the
gas pressure does not change much across the prominence. According to the ideal
gas law a decrease in density leads to an increase in temperature to keep
pressure constant. 

We have repeated the simulation for the configurations with intermediate and weak
twist, and the results for the three twisted configurations at the same instant
($63.6\,\rm min$) are shown in Figure~\ref{prom2}. For the three configurations we obtain
suspended prominences, and this is not surprising for the cases with strong and
moderate twist since the magnetic structure has dips. Nevertheless, even in the
situation with weak twist we obtain a suspended prominence although there are no
magnetic dips. The mass deos not fall along the curved magnetic field lines because an
overpressure is established between the prominence and the base of the corona. This
overpressure inevitably produces an increase in temperature in the coronal medium,
which is at most $10\%$ of the coronal temperature ($1\,\rm MK$) but radiative losses
and conduction, not present in our model, would try to reduce this temperature
increase. This particular support of the prominence is the direct consequence of the
use of line-tying conditions at the bottom boundary because pressure perturbations are
not allowed to leave the system. In the low-$\beta$ regime gas pressure changes are
strongly localized along the magnetic field lines, and the prominence behaves like a
piston that moves downwards until it reaches an equilibrium state.

For the spatial resolution used in the previous simulations ($600\,\rm km$), the
configurations for moderate and weak twist shown in Figure~\ref{prom2} are able to
get to a situation which is close to a quasi-stationary state. For these two
cases the newly achieved configuration could be used as a background equilibrium
in studies of for example, transverse or longitudinal oscillations.

\subsection{Time evolution of CM}

A detailed analysis of the changes in the prominence reveals that it is in fact
quite dynamic, involving rather complicated motions before the quasi-stationary
situation is reached.  In order to quantify the   three-dimensional movement of
the prominence its center of mass (CM) has been calculated. The computation of this
magnitude has been done using the three-dimensional information of the density in a
box that includes the prominence body only, avoiding a significant contribution from
the background stratified atmosphere present in the whole computational domain. The
calculations have been performed using the visualization tool VisIt \citep[][]{
VisIt}

\begin{figure}[!ht] \center{\includegraphics[width=8.cm]{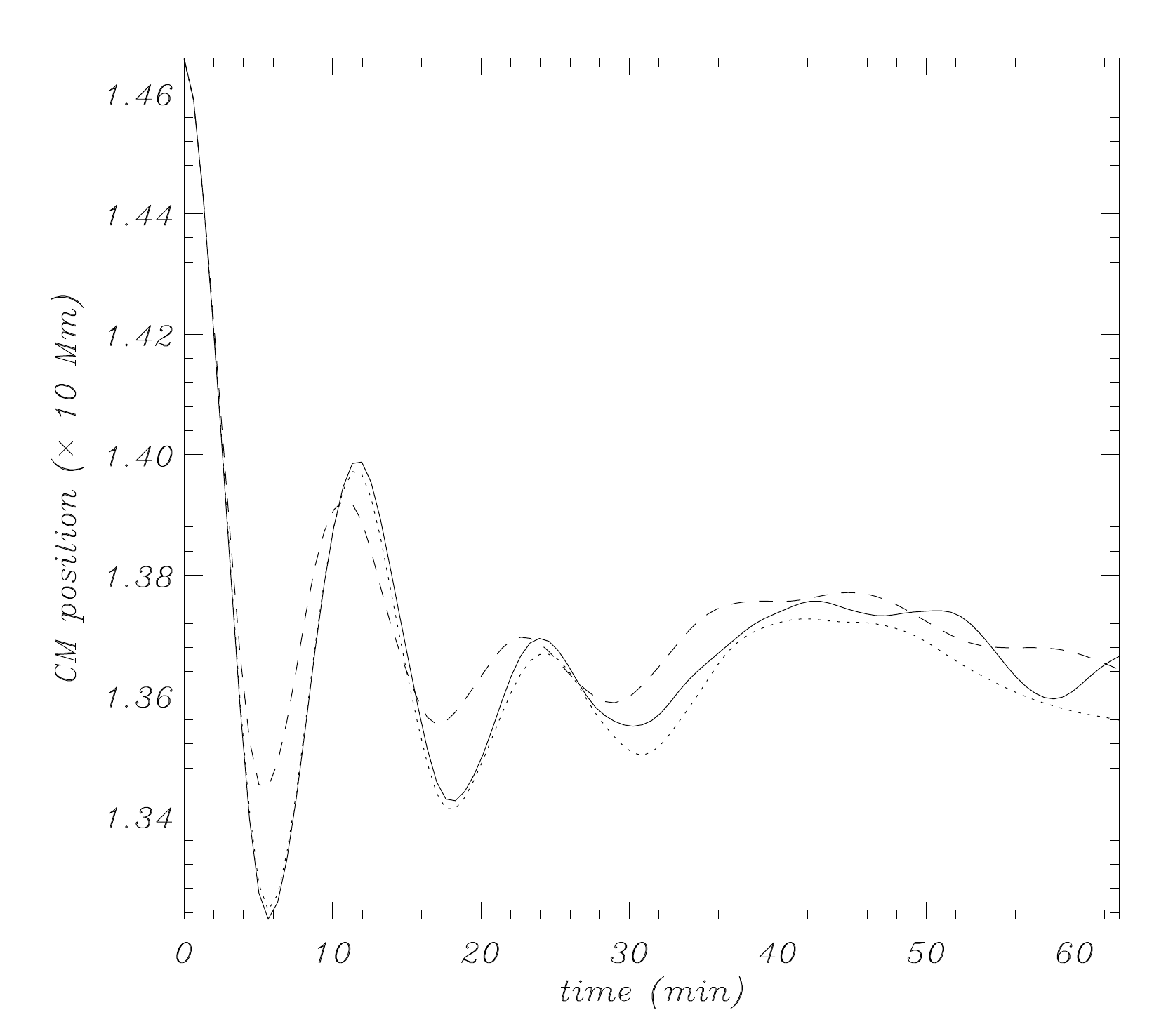}}
\caption{\small Position of the center of mass for three different grid
resolutions, $1200\,\rm km$ (dashed line), $600\,\rm km$ (dotted line) and
$300\,\rm km$ (continuous line). For this case $N_t=8.5$.}\label{comparres}
\end{figure}

Due to the symmetry of the system, the $x$ and $y-$coordinates of the CM are equal to
zero, while the $z-$coordinate changes with time. The evolution of this component is
represented in Figure~\ref{comparres}, and provides a proxy for the global vertical
motion of the prominence. In this plot, see continuous line, we can distinguish several
oscillations. Initially the prominence descends but around $t=6\,\rm min$ it starts to
displace in the opposite direction indicating an upward motion. Later the direction of
the motion is reversed several times. This oscillatory pattern lasts for about two
periods, i.e., $35\,\rm min$. The characteristic period is, for this simulation, around
$12\,\rm min$.  The periodicity of the CM reflects a global vertical oscillation of the
prominence.

\begin{figure}[!ht] \center{\includegraphics[width=8.cm]{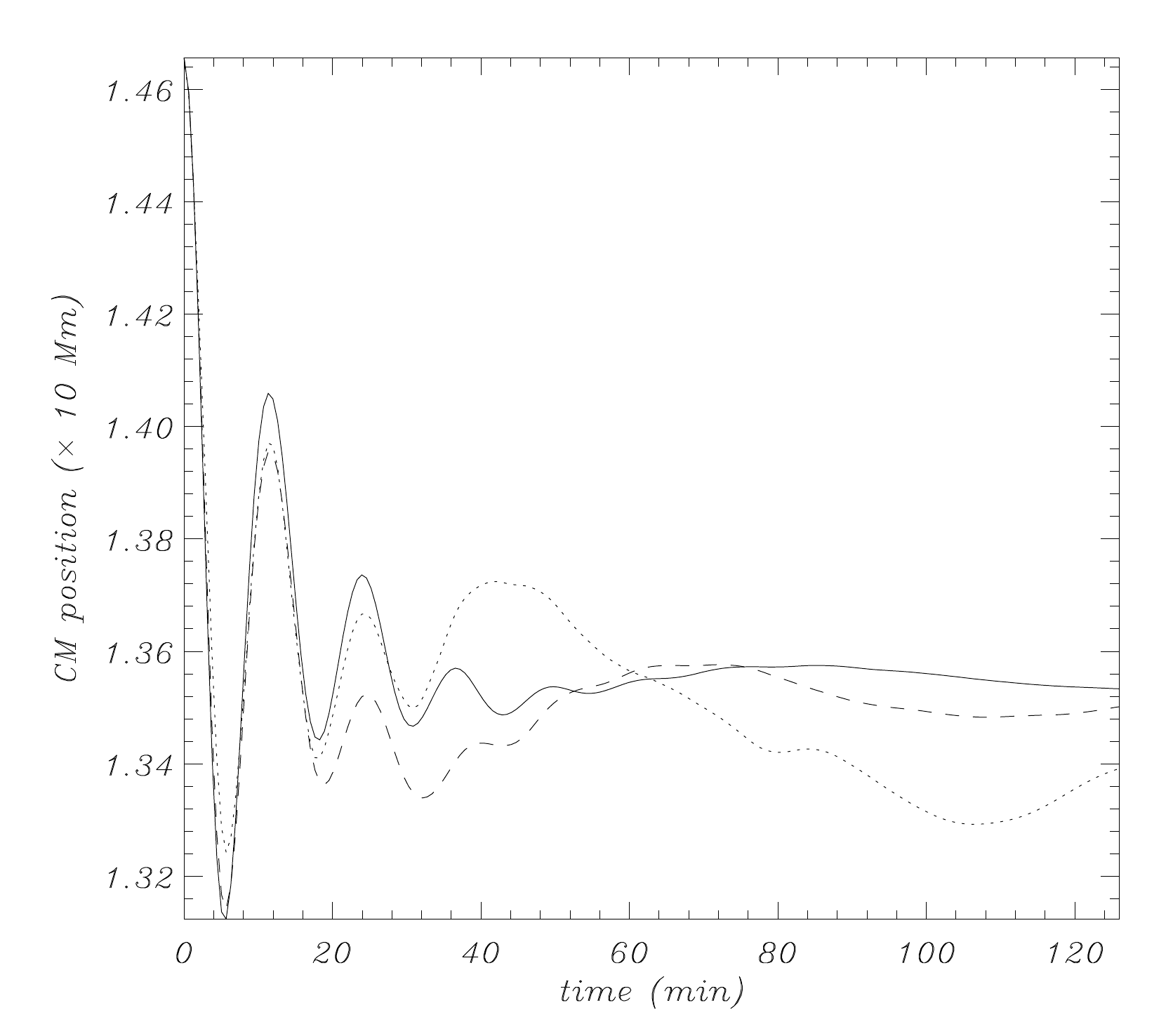}}
\caption{\small Position of the center of mass for three different values of the
twist, $N_t=8.5$ (dotted line), 4.3 (dashed line), 2.2 (continuous line). In
these simulations the resolution is $600\, \rm km$.}\label{compartwist}
\end{figure}

Figure~\ref{compartwist} shows the position of the CM as a function of time for the
different values of twist but for a longer run (about two hours). The structure tends to
settle down at essentially the same height, around $13.5\,\rm Mm$ for weak and moderate
twists. For the case with strong twist the prominence still experiencies significant
changes after one hour of evolution. Surprisingly, the periodicities during the initial
relaxation process, from $t=0$ to $t=40\,\rm min$, are very similar for the three
configurations, meaning that twist does not affect significantly the period of oscillation
of the global standing transverse mode. This is in agreement with the eigenmode results of
standing transverse kink oscillations in straight cylindrical tubes with weak magnetic twist
\citep[see][]{ruderman07,terradasgoossens2012,ruderterr2015}. Furthermore, the behavior
found for  $N_t=2.2$ and 4.3 after $60\,\rm min$ of evolution reveals that these
configurations are close to a quasi-stationary state.

Interestingly, for the three values of magnetic twist the attenuation of the CM height with
time (in the range $t=0$ to $t=40\,\rm min$) is quite strong. The attenuation or damping is
a topic that has been discussed extensively, for example, in the context of coronal loop
oscillations. In particular the attenuation in ideal MHD can be associated to the process of
resonant absorption and/or  wave leakage. Nevertheless, dissipation due to the numerical
scheme used in the codes can be also a source of dissipation that is very often disregarded.
In order to quantify the potential effect in our simulations and in particular on the
attenuation of the position of the CM, we have performed a convergence test, and have
calculated the CM height for three different grid resolutions. According to
Figure~\ref{comparres}, convergence of the results is achieved for resolutions between
$600\,\rm km$ (dotted line) and $300\,\rm km$ (continuous line), since the two corresponding
curves are not significantly different. The attenuation is not highly affected by the
resolution as long as we use a grid separation smaller than $600\,\rm km$. A resolution of
$1200\,\rm km$ leads to shorter periods and faster attenuation of the global displacement
and is not suitable to perform quantitative studies. We can, therefore, claim that the
attenuation of the CM position is real (physical) and not dominated by numerical dissipation. This
does not mean that at a certain time in the evolution numerical dissipation can be relevant.
This issue is discussed in the following sections.

\begin{figure}[!ht] \center{\includegraphics[width=8.cm]{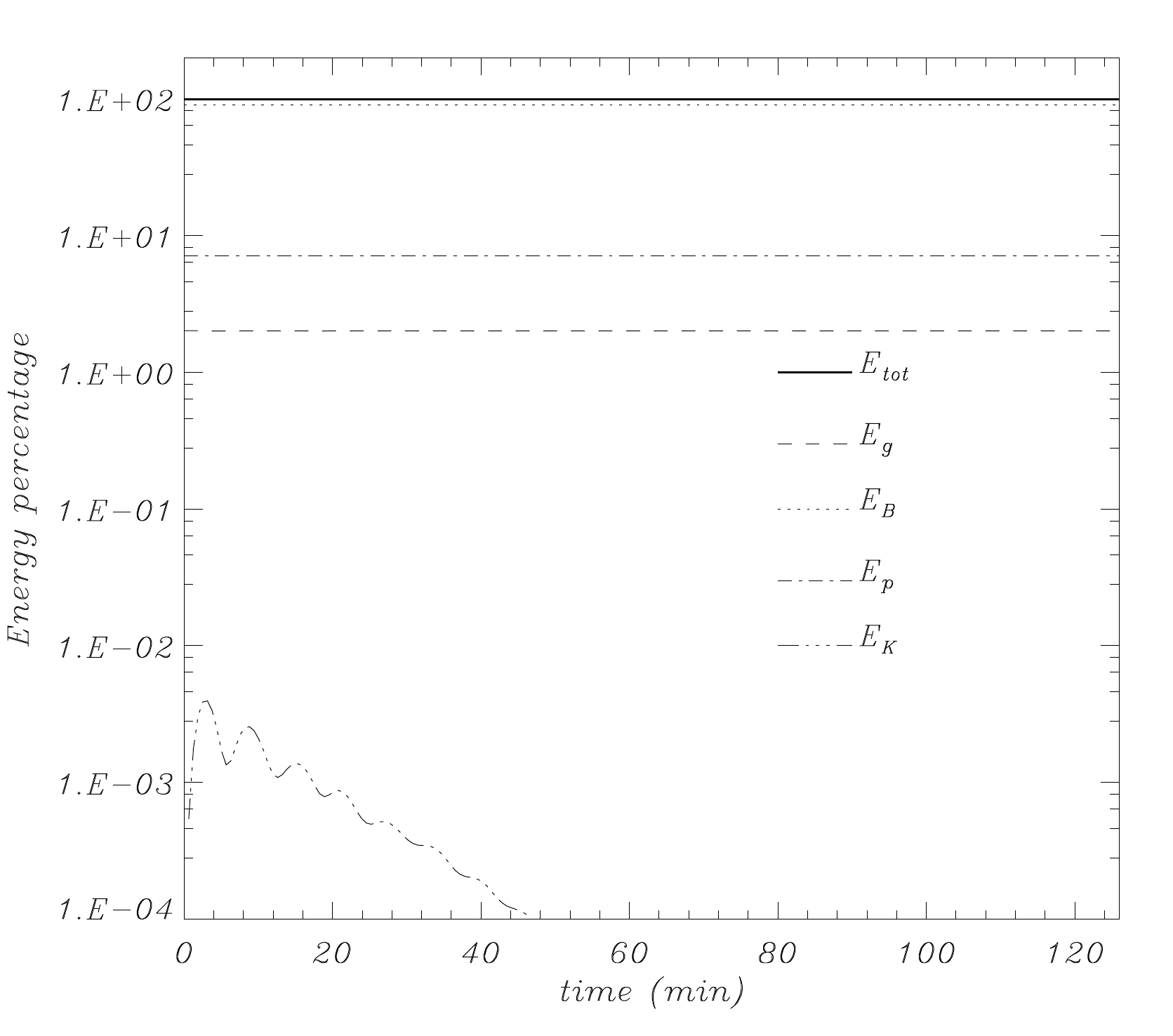}}
\caption{\small Total energies as a function of time. Note the logarithmic scale
in the vertical axis. In this simulation the resolution is $600\, \rm km$ and
$N_t=2.2$.}\label{energt} \end{figure}

\section{Energy considerations and damping}

\subsection{Closed boundaries}

To clarify the origin of the attenuation of the CM position, the energy of the system is
investigated. Closed boundary conditions (see Section~\ref{method}) are first
considered (the simulations described so far made use of these conditions). This
means that energy is unable to escape from computational box, and ideally the
total energy of the full system should be conserved since the governing
equations do not include explicitly dissipative mechanisms. However, numerical
dissipation can produce an energy loss. In Figure~\ref{energt} the different
contributions to the total energy are represented as a function of time for the
most simple configuration ($N_t=2.2$). These energies are integrated over the
whole 3D domain and have been normalized to the total energy at $t=0$. From the
plot we find that since the plasma-$\beta$ is low, the magnetic energy ($E_B$)
has the largest contribution, more than 90\% of the total energy.  The internal
energy ($E_p$) represents about 8\% while the gravitational ($E_g$)  2\%.
The kinetic energy ($E_K$) is rather small in comparison with the total energy
budget ($E_{tot}=E_B+E_p+E_g+E_K$) of the system. Nevertheless, this kinetic
energy accounts for all the motions in the prominence.

\begin{figure}[!ht] \center{\includegraphics[width=8.cm]{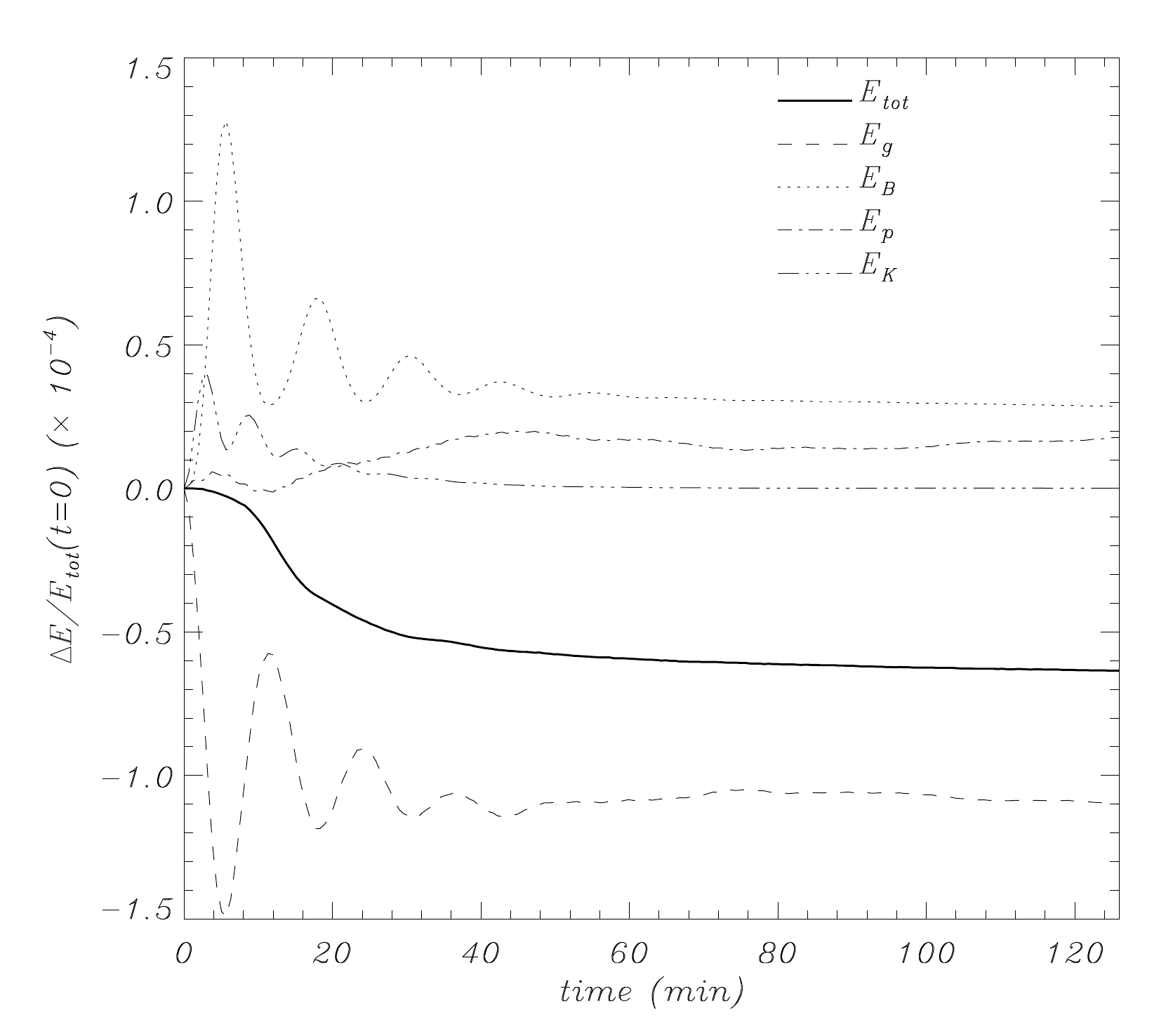}} \caption{\small
Percentage of differences of energies with respect to theri values at $t=0$ as a
function of time. Based on the results of Figure~\ref{energt}.}\label{energ0}
\end{figure}

\begin{figure}[!ht] \center{\includegraphics[width=8.cm]{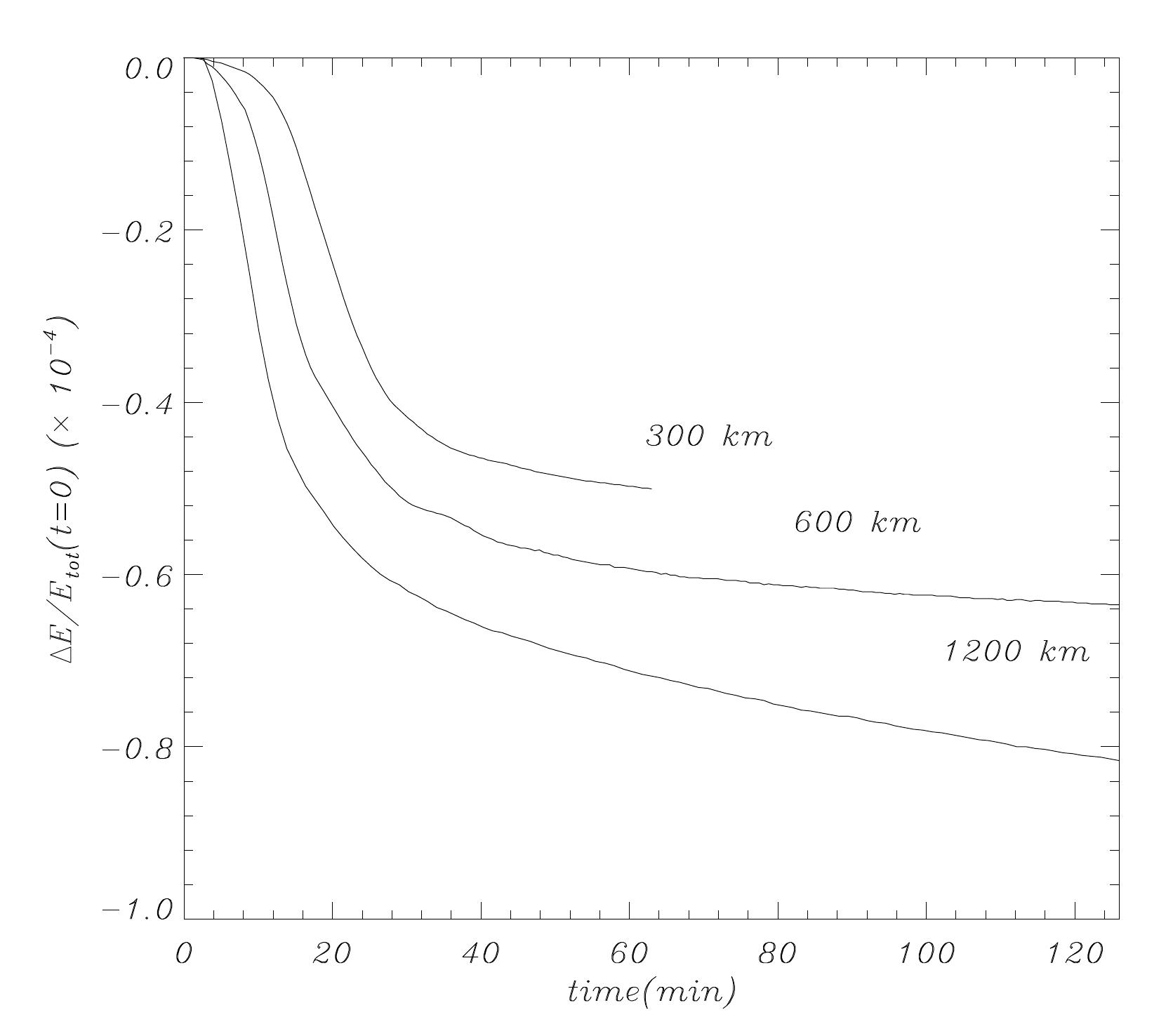}}
\caption{\small Change in total energy for different grid resolutions. In this
simulation $N_t=2.2$.}\label{energ1} \end{figure}

From Figure~\ref{energt} it seems that the total energy is constant with time
because of the scale used in the vertical axis of the plot. A more elaborated
analysis reveals that indeed part of the total energy is lost. In 
Figure~\ref{energ0} we have now represented the energy differences with respect to
the initial state normalized to the total energy at $t=0$, allowing a better
interpretation of the results. The magnetic and internal energies for large
times increase with respect to the initial value while the gravitational energy
decreases. This agrees with the global motions described before, the prominence
position decreases with time while it oscillates and settles at a lower height,
compressing the magnetic field and increasing the gas pressure. The kinetic
energy initially increases (when the prominence starts to move downwards) but
soon it decreases until it reaches very low values. More important, from the
curve of the total energy differences we realize that after around $t=10\, \rm
min$, there is a sudden decrease followed by a quite stationary situation. This
is a consequence of the development of small scales below the spatial grid
resolution and the effect of numerical dissipation. This is corroborated by
performing the same plot for different grid resolutions, see Figure~\ref{energ1}.
The better the resolution the later the decrease in energy since smaller scales
are resolved.  The curves in  Figure~\ref{energ1} do not converge to the same
value. This is because when the spatial resolution is increased the overall
numerical dissipation is  smaller, meaning that globally less energy is
dissipated. Ideally one would expect the change in energy to go to zero as the
grid resolution is increased. 

It is necessary to remark that this energy loss does not affect significantly
the damping time since we have already demonstrated that increasing the
resolution gives essentially the same attenuation (see Figure~\ref{comparres}).  
This means that the attenuation in our system is not related to the numerical
energy loss and it has a physical origin. Note also how the oscillations are
very clear in Figure~\ref{energ0} and for the kinetic energy have a period that is
half of that of the vertical oscillation since $E_K$ varies quadratically with
the velocity.

\begin{figure}[!ht] \center{\includegraphics[width=8.cm]{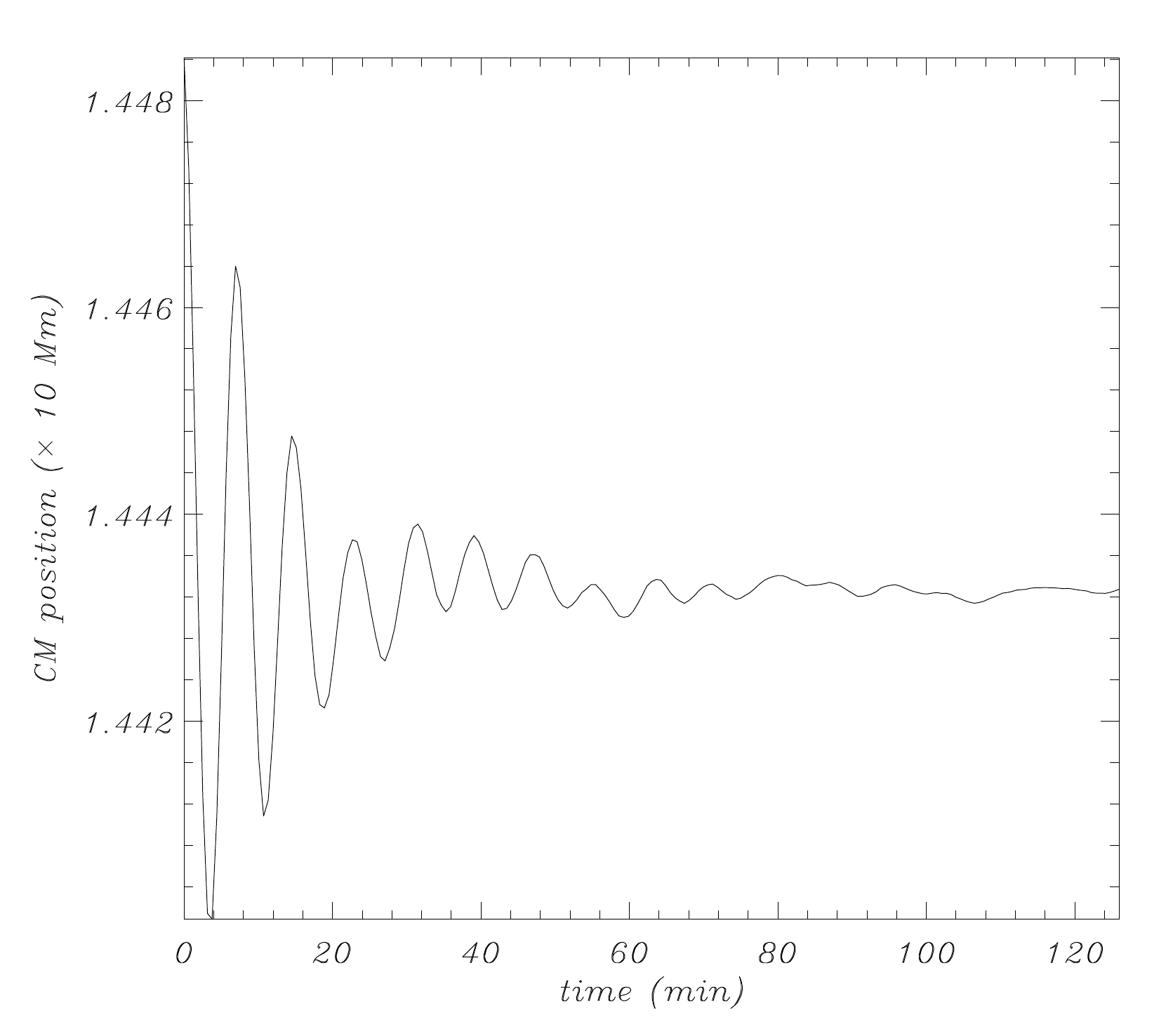}}
\caption{\small Position of the center of mass for the case with a low
prominence-corona density contrast. For this case $N_t=2.2$, and the resolution
is $600\,\rm km$.}\label{cmlight} \end{figure}

With the aim of understanding the physical origin of the attenuation of the CM
position with time, we have performed a simulation for a prominence with a very low
density contrast. In this specific simulation we chose a value of 5 for the density
enhancement, which is certainly unrealistic but it will shed light on the damping
mechanism. The reason for decreasing the density contrast is that this allows us to
study a prominence with a very small mass. Such a light prominence will not
experience large motions in the vertical direction in comparison with the previous
simulations, and hence nonlinear effects, discussed in the following Section, are
expected to be very weak. The results of the simulation are shown in
Figure~\ref{cmlight}. The basic period of oscillation, around $8\,\rm  min$, is
shorter than in Figure~\ref{compartwist} (because the prominence is lighter), and
the attenuation is weaker (now we can count more than 9 periods). There is also
evidence in Figure~\ref{cmlight} of a longer period, which is around $43\,\rm min$.
The physical meaning of this periodicity is related with a second motion of the
structure of different origin from the vertical oscillation. It corresponds to a
global motion but along the magnetic field, and due to the symmetry in the system
this oscillation is antisymmetric with respect to the center of the prominence,
meaning that it produces contractions and dilatations of the whole prominence. In
terms of MHD waves it corresponds to a slow antisymmetric mode, and this explains
the long periodicity since slow modes are essentially driven by pressure forces,
which in our case are rather weak (low-$\beta$ regime). This mode of oscillation is
also responsible for the long periodicities found in  Figure~\ref{compartwist} for
times longer than $40\,\rm min$, referred before as the quasi-stationary situation.
In fact this periodicity is present from $t=0$  and is superimposed to the shorter
periodicity associated with the global vertical oscillation. 

\begin{figure}[!ht] \center{\includegraphics[width=7.cm]{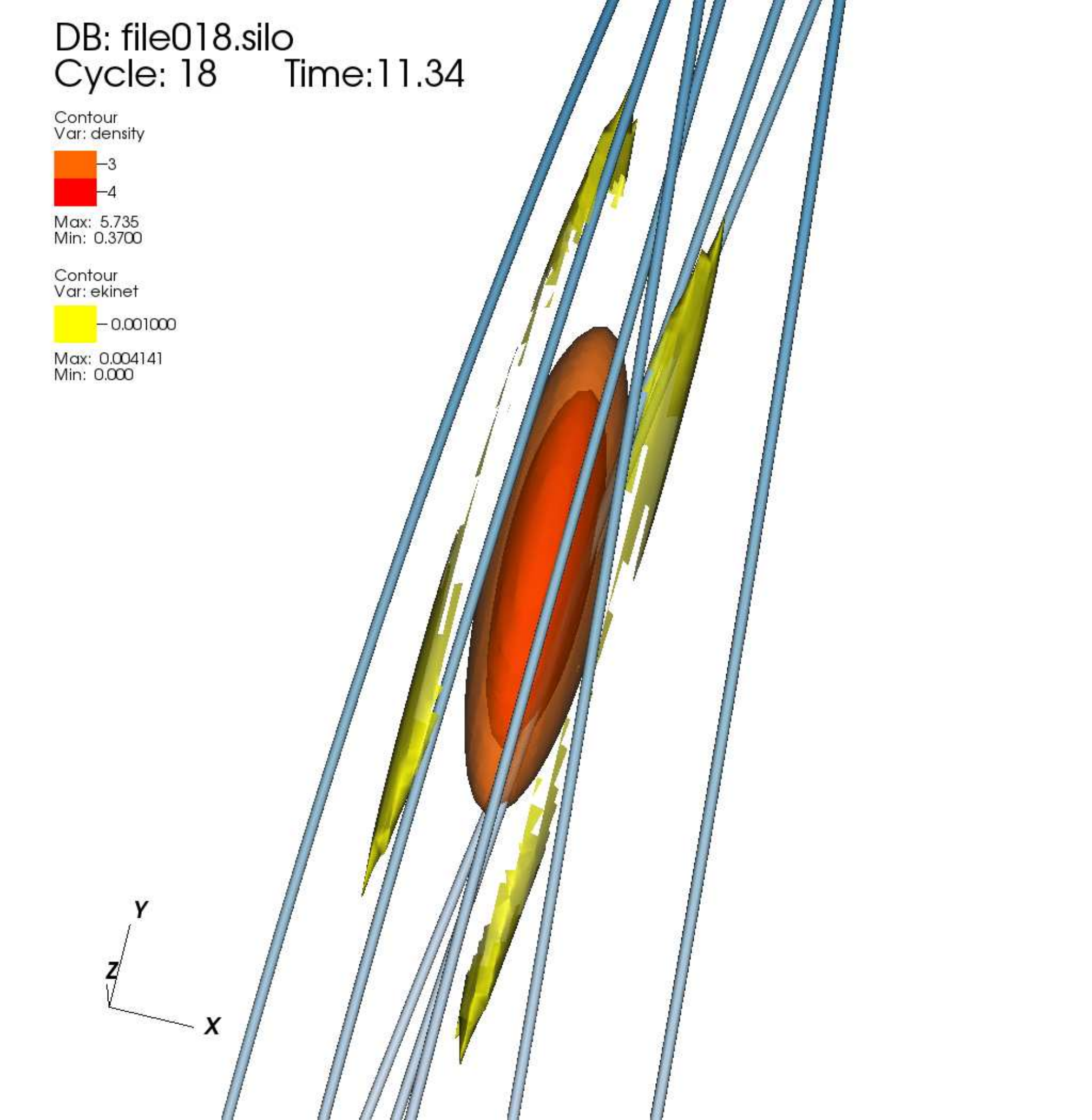}}
\center{\includegraphics[width=7.cm]{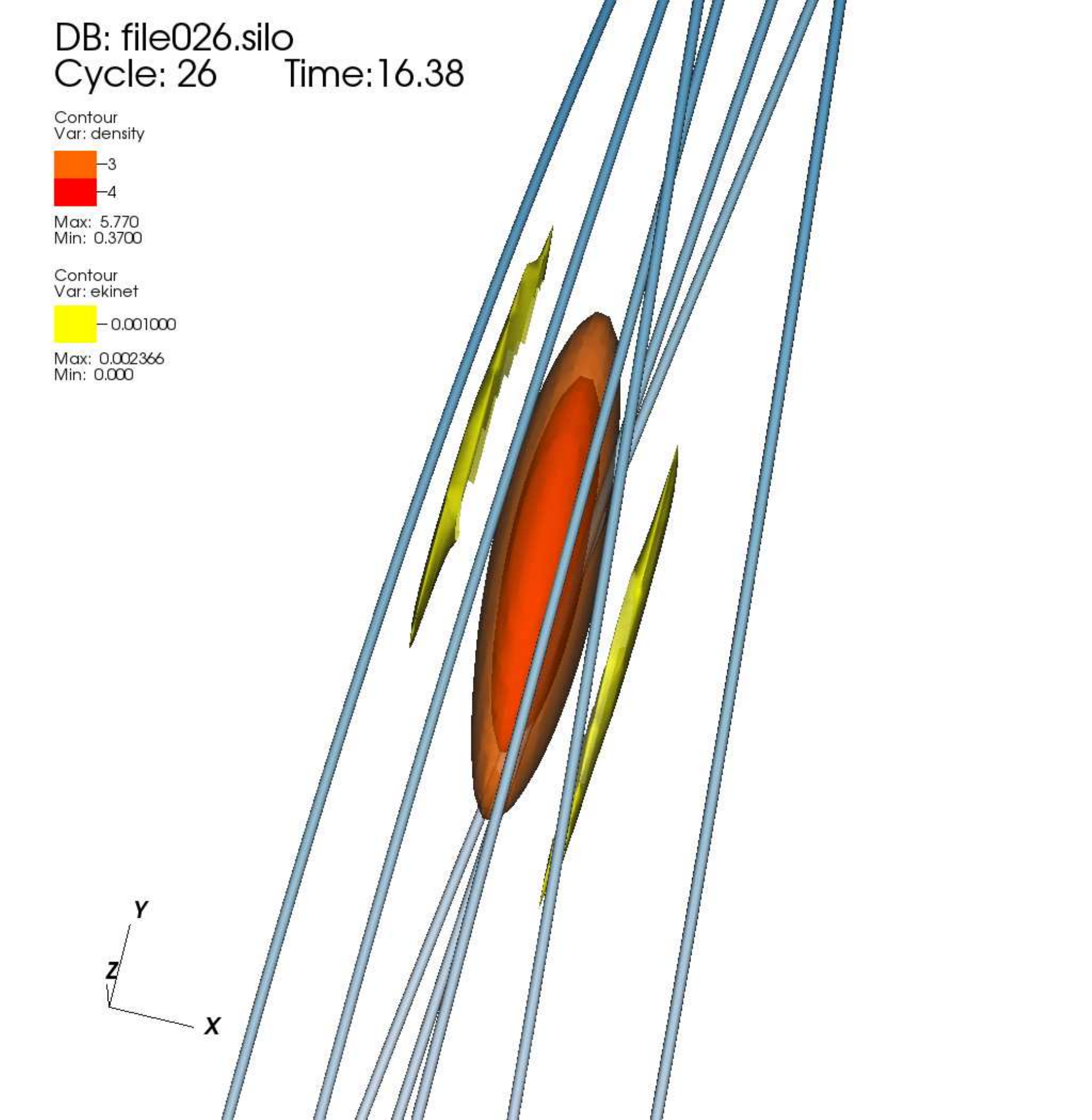}} \caption{\small Detail of
the prominence (red-orange colors) with some magnetic field lines and the locations
where the kinetic energy increases (yellow colors) due to continuum damping. For
this case $N_t=2.2$. See movie in the online version of the journal.}\label{reslayer}
\end{figure}

\begin{figure}[!ht] \center{\includegraphics[width=8.cm]{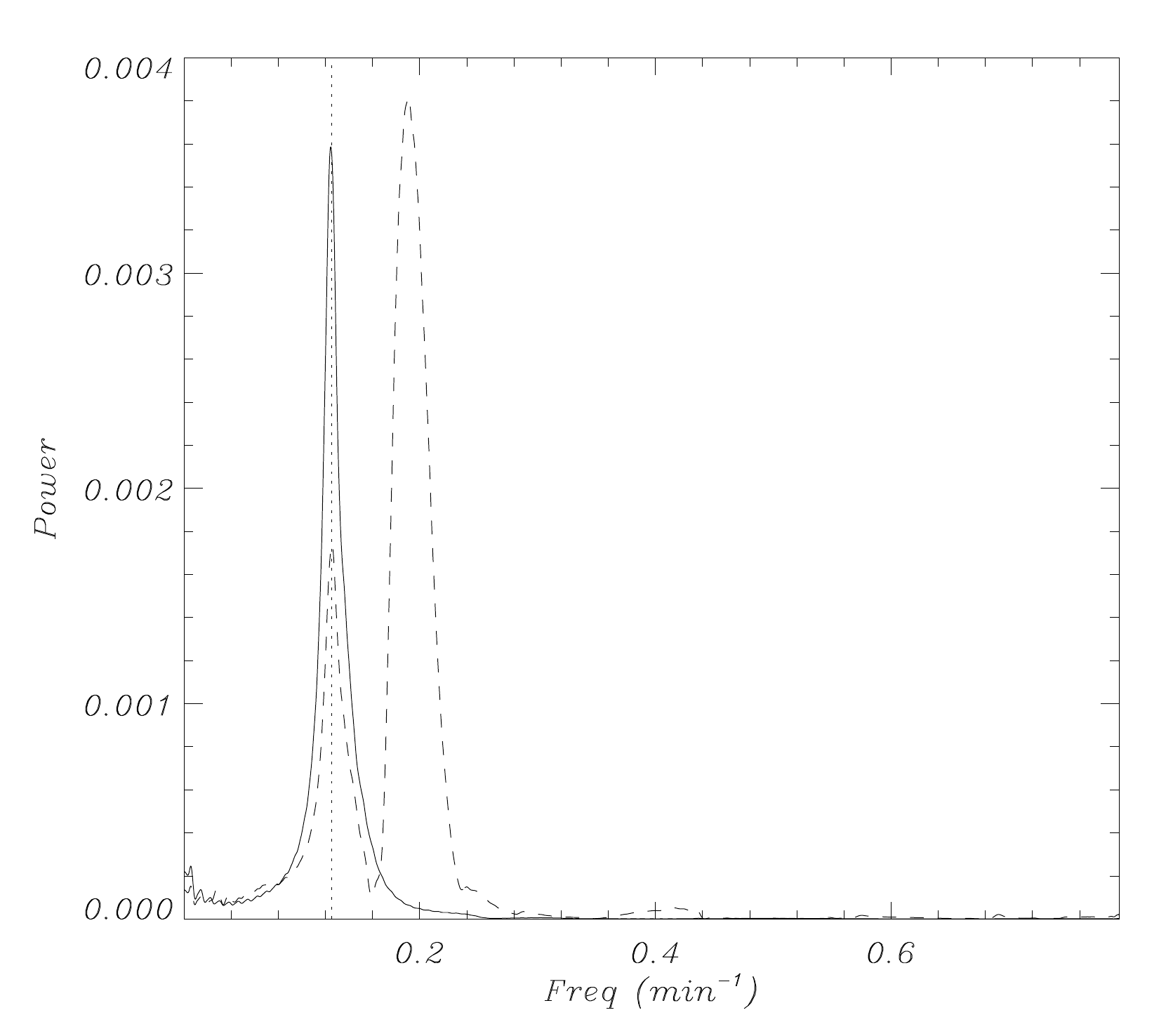}} \caption{\small
Power spectrum of the vertical velocity component at two different positions. The
continuous line corresponds to the point close to the center of the prominence while
the dashed line is in the PCTR. The two power spectra show a common peak (plotted
with a vertical  dotted line), associated to the global motion of the prominence
(quasi-mode).}\label{periodogram} \end{figure}

We return to the issue of the damping of the vertical oscillation. A
comprehensive analysis of the results for the light prominence indicates that
the reason for the attenuation is due to the conversion of energy of the
transverse vertical motion into localized motions at the lateral edge of the
prominence. This is equivalent to the continuum damping mechanism that has
been investigated in simpler configurations (such as plasma slabs or magnetic
cylinders). In our simulations, although the geometry is complicated, one can
identify the locations where there is an increase in kinetic energy due to this process.
In Figure~\ref{reslayer} density isocontours, representative of the shape of the
prominence, are plotted together with a specific isocontour of kinetic energy.
It is in a thin region near the edges of the prominence, essentially at the
PCTR,  where the energy is transferred and concentrated. The power spectrum of
the vertical velocity component calculated at different locations provides additional
information relevant to the resonant absorption process. Two points localized at
the central plane ($x=0$), one close to the center of the prominence (point 1)  and
the other at the PCTR (point 2), have been selected. The corresponding power
spectra are plotted in Figure~\ref{periodogram}. At point 1 the signal has a
dominant periodicity of $8\,\rm min$. For point 2, two peaks are clearly
discernible, one located around $8\,\rm min$ and the other peak is around 5.25 min. Since the two points share a common period we associate this
periodicity to the global mode. This period agrees with the one found in
Figure~\ref{cmlight} and corresponds to the global vertical oscillation of the
prominence. For point 2, the secondary peak corresponds to a particular Alfv\'en
mode that belongs to the Alfv\'en continuum. If we chose another point in the
PCTR the frequency spectrum shows power at a different Alfv\'en frequency. \citet{terrarretal08} already found this behavior in a rather complicated
multistranded loop geometry. In their Figure~3 the signal at a given point shows
the collective or global frequency and also the local Alfv\'en frequency. The
amplitude of the global mode decreases with time while the amplitude of the
local Alfv\'en modes increases with time due to the energy transference. 

The excitation of local Alfv\'en modes is also clear in driven problems. For example
\citet{wrigth1992} using a simple model demonstrated how under different driving conditions,
resonant and nonresonat Alfv\'en modes can be excited in the system \citep[see
also][]{degroofetal02,degroofgoossens02}. In fact, some of the results of our simulations
can be understood assuming that the global mode acts as a driver at a given frequency of
finite duration. \citet{wrigth1992} showed that for a magnetic field line where the
continuum frequency  is different to the driving frequency the field line responds at two
frequencies: The natural Alfv\'en frequency and the driven frequency. This is exactly what
Figure~\ref{periodogram} shows for point 2. In contrast, for a magnetic field line close to
the center of the prominence, like for point 2, the frequency of the driver is essentially
the same as the natural Alfv\'en frequency and that is why the power spectrum shows a single
peak.

To summarize, in the time-dependent problem studied in our simulations, the initial
global oscillation of the whole prominence is transferred to the Alfv\'en continuum modes of
the PCTR providing compelling evidence of the presence of continuum damping. This energy
transference is most efficient where the frequency of the global mode matches the local
Alfv\'en frequency, but there is also energy transference to neighboring field lines that
have their own Alfv\'en frequency. The amplitude of the Alfv\'en continuum modes grows with
time until all the energy of the global mode has been transferred. Subsequently the
continuum modes oscillate with their natural frequencies. Since the frequency of the
Alfv\'en modes changes with position this enhances the phase-mixing process. Small spatial
scales are generated and eventually smoothed by the numerical dissipation of the scheme.

\subsection{Open boundaries}

Now open boundary conditions are considered, allowing the role of energy leakage and its contribution to
the observed attenuation  of the position of the CM to be investigated. The results indicate that for
the present configuration the energy loss due to wave leakage through the boundaries is rather small
since the position of the center of mass is essentially the same as in the situation for closed
boundaries. Thus wave leakage is not the main cause of damping of the vertical oscillations. The
dominant physical process is transfer of energy to Alfv\'en oscillations of
neighboring field lines.

There is additional indirect evidence about the absence of significant leakage in
our system. From the calculation of the power spectrum at different points, described
before, and the inferred period of the global oscillation, wave leakage of this mode is
not possible from the theoretical point of view. The cause is that the Alfv\'enic modes
in the coronal environment outside the prominence have frequencies that are above the
frequency of the global mode. This is incompatible with the properties of leaky modes,
that are fast magnetoacoustic modes with frequencies above the local Alfv\'en
frequency, and therefore have a propagating nature. This propagating feature is
precisely the responsible for the attenuation of the mode since the energy of the
global mode is emitted away. These types of modes are not possible in our
configuration.

\begin{figure}[!ht] \center{\includegraphics[width=7.cm]{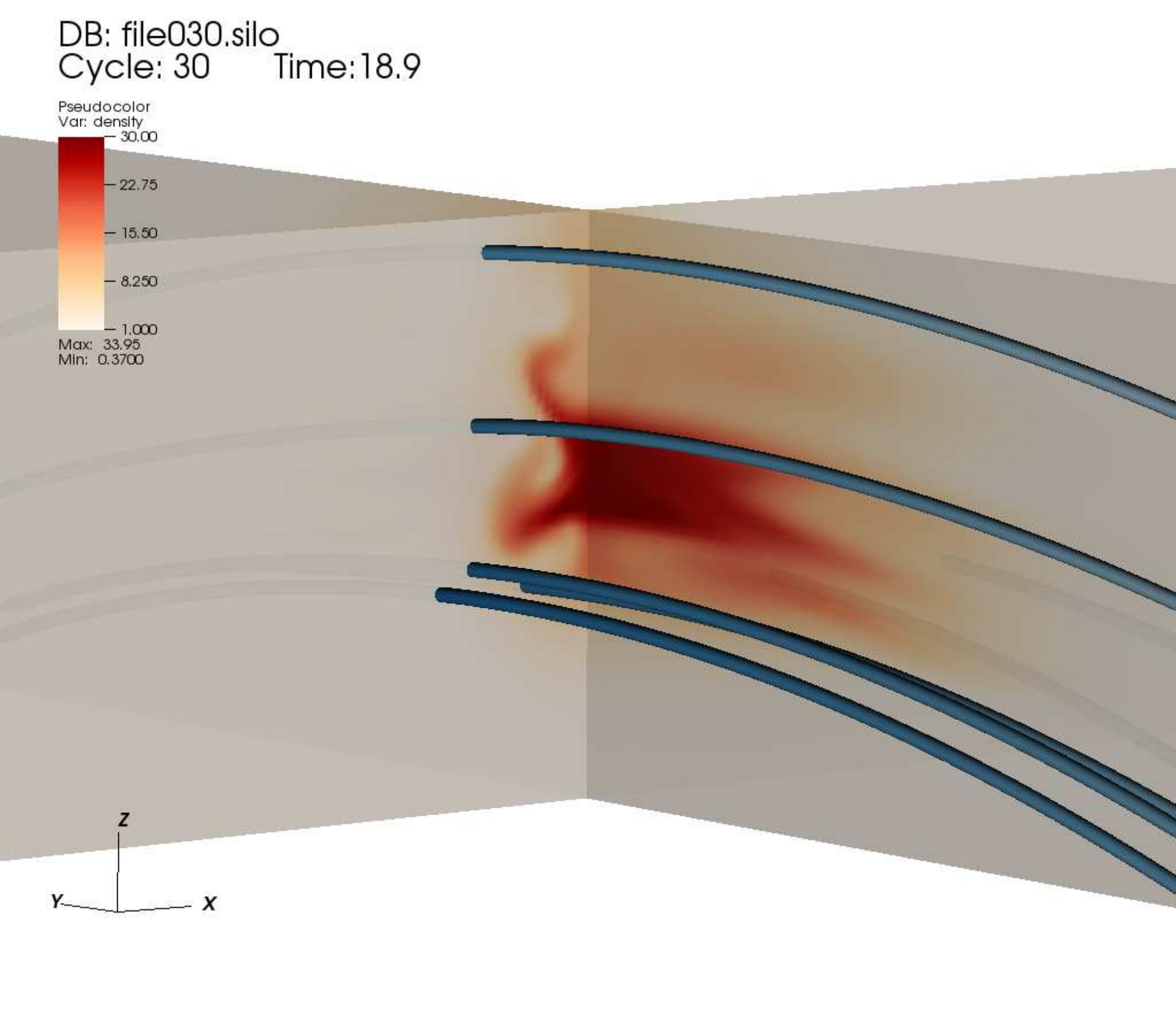}}
\center{\includegraphics[width=7.cm]{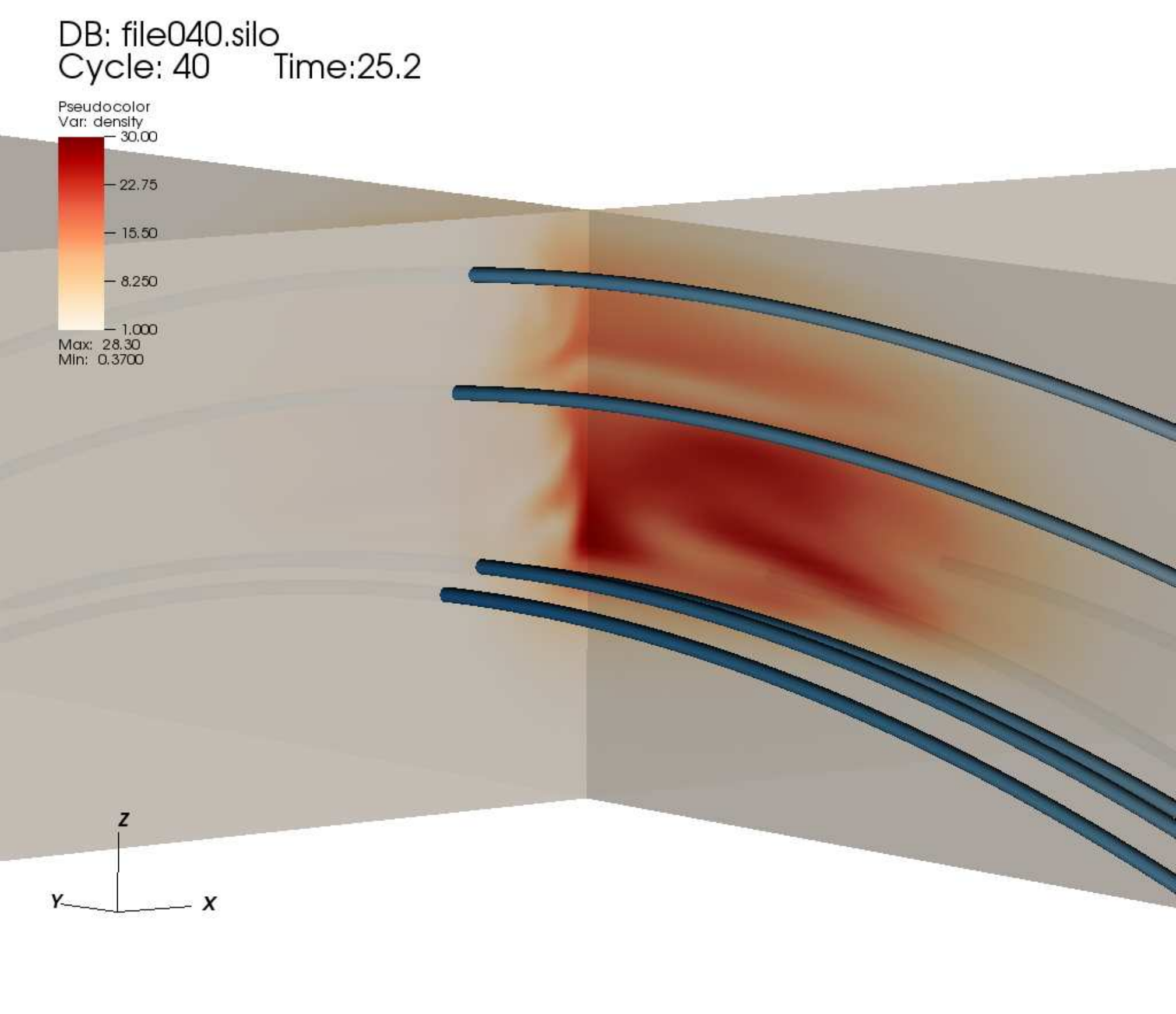}} \caption{\small Detail
of the density evolution at two perpendicular planes. For this case $N_t=2.2$
and the spatial resolution is $300\,\rm km$. }\label{khnt2_2} \end{figure}

\begin{figure}[!ht] \center{\includegraphics[width=7.cm]{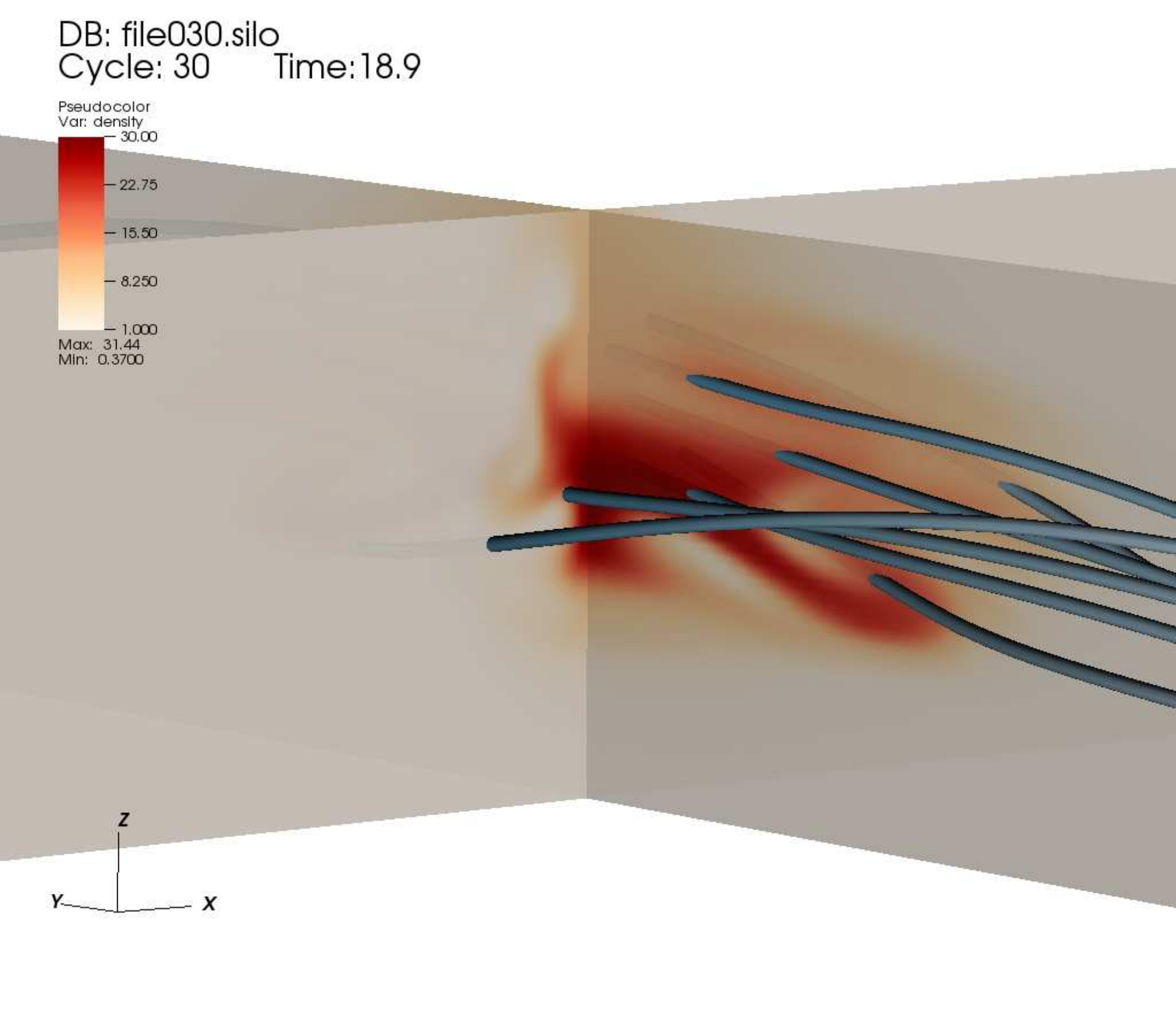}}
\center{\includegraphics[width=7.cm]{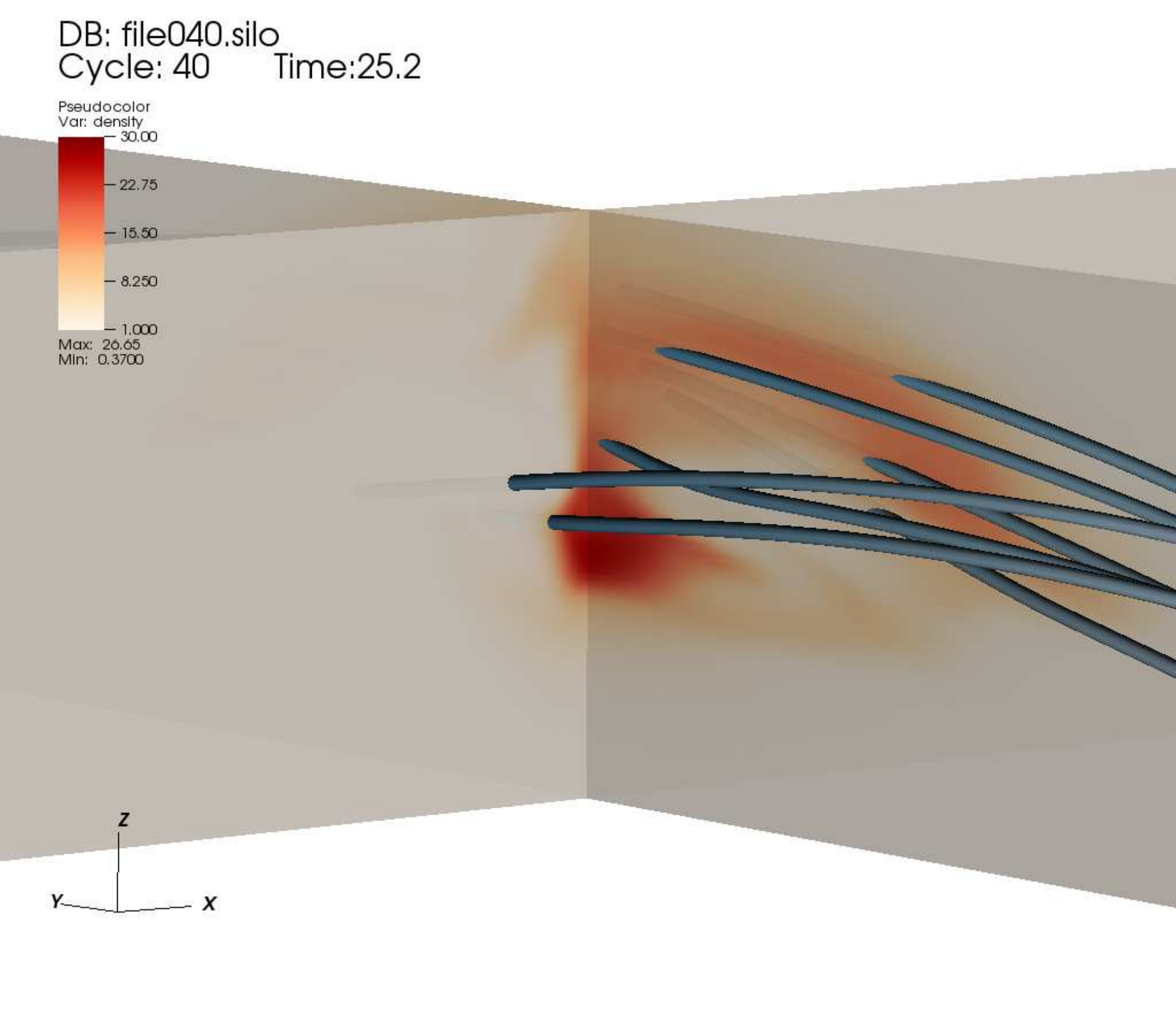}}\caption{\small Same as in 
Figure~\ref{khnt2_2} but for $N_t=4.3$. }\label{khnt4_3} \end{figure}

\section{Kelvin-Helmholtz instability}\label{kh}

The results of the previous Section indicate that resonant absorption is responsible for the attenuation
of the position of the CM. This is true at least for the case of the light prominence. The results for
the typical prominence considered in our simulations (see  Figure~\ref{compartwist}) also suggest a strong
damping. However, in this last case the physics involved in the attenuation is more intricate, specially
due to nonlinear phenomena. The reason is that since now the prominence is more massive, the vertical
motions and therefore the shear motions at the PCTR, have larger amplitudes and ultimately produce a KHI
type.  This instability operates in the absence of inhomogeneous layers, i.e., for a discontinuous
change in the velocity, but also under the process of resonant absorption and the associated
phase-mixing. \citet[][]{terradasetal08,antolinetal2015} have studied numerically this KHI using
straight cylinders with nonhomogeneous layers in the radial direction, while \citet{soleretal2010} have
performed an analytic study assuming a jump in the velocity shear. The flux rope configuration studied
in the present work, although is more complex than the simple straight tube, is not an exception. The
development of the instability is quite evident in the simulations with the highest resolution. For the
intermediate spatial resolution ($600\,\rm km$) it is not so obvious. An example with a resolution of
$300\,\rm km$ is shown in Figure~\ref{khnt2_2} for the case $N_t=2.2$. In this plot the density
distribution at two perpendicular planes that intersect at the center of the prominence are plotted
together with some selected magnetic field lines. The initially compact shape of prominence progresses
and ripples at the sides of the prominence start to appear (see top panel). These are precisely the
locations where the velocity shear is strong. Later, small scales are still developing and some
structuring is devised along the field lines (see bottom panel).

\begin{figure}[!ht] \center{\includegraphics[width=7.cm]{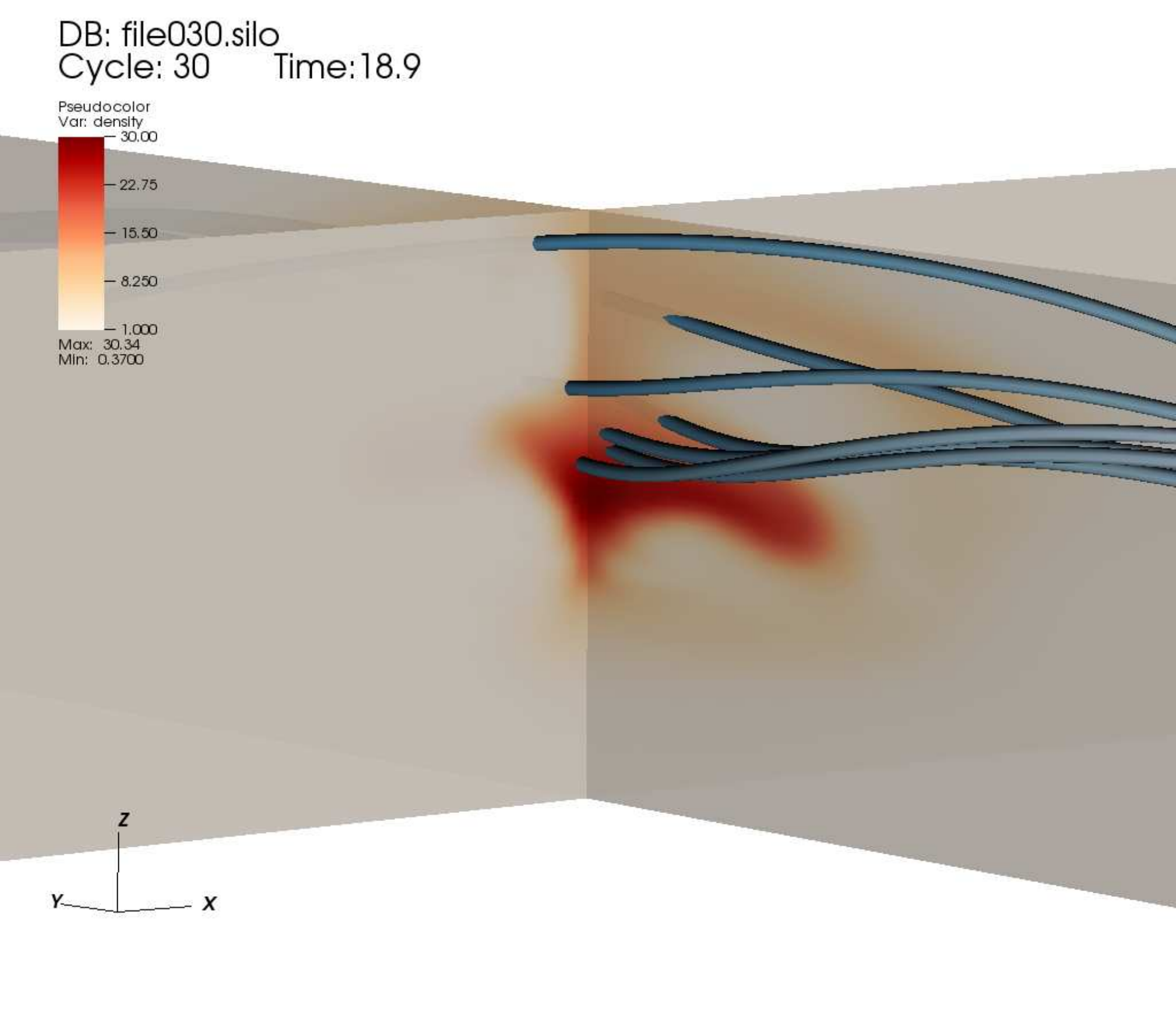}}
\center{\includegraphics[width=7.cm]{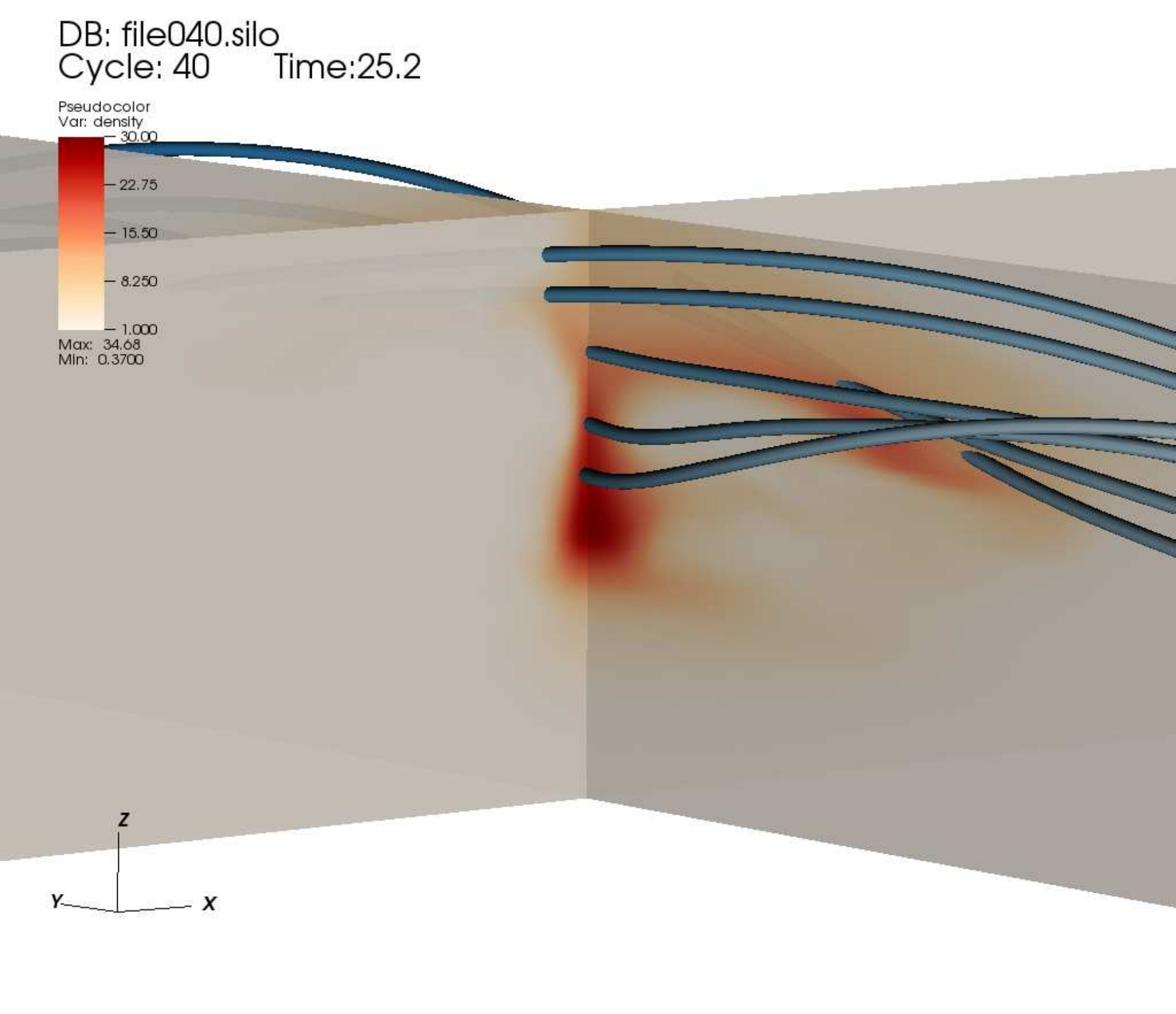}} \caption{\small Same as in 
Figure~\ref{khnt2_2} but for  $N_t=8.5$. }\label{khnt8_5} \end{figure}

A question that arises here is whether the attenuation, discussed in the previous Section
for a light prominence, is produced by resonant absorption at the inhomogeneous layers, or
if it is instead more related to the nonlinear effects that generate the vortical structures
at the sides of the prominence due to the KHI. In other words, the KHI by itself may provide
a mechanism to attenuate the prominence even in the absence of resonant absorption since
this  instability precisely takes energy from the bulk flow (the transverse vertical
oscillation in our case) that eventually cascades to smaller scales. Depending on the
amplitude of the oscillation (the KHI is a nonlinear problem while resonant damping works
linearly) the KHI could transfer the energy more efficiently than
resonant absorption. This is an attractive problem that requires a careful analysis,
preferably using a simpler model like the straight cylindrical tube, and is left for
future studies. In this respect, \citet{rudermanetal2010} have proposed that the
nonlinear coupling between kink and fluting modes may accelerate the damping of the
oscillations since fluting modes have a faster attenuation rate.

Finally, it is interesting to investigate the effect of twist on the development of the KH instability
for the highest spatial resolution. In Figs.~\ref{khnt4_3} and~\ref{khnt8_5}, the density distribution
for the configurations with moderate and strong twist are plotted. We see that the ripples associated
with the instability develop in a different way, for moderate twist they are clearly visible at
intermediate times (top panel of  Figure~\ref{khnt4_3}). These vortical structures are not so evident in
Figure~\ref{khnt8_5}, representing the situation with the strongest twist, but a detailed inspection of
the motions shows that  the instability is still present. It is known that magnetic twist must have a
stabilizing effect \citep[see][]{soleretal2010}, but in the present configuration, even for the
situation named here as strong twist, it is still too weak to inhibit completely the development of the
instability. A distinctive feature of the flux rope with weak twist, showing quite a compact shape, in
comparison with the situations of moderate and strong twist is that in these last two cases elongated
structures along the flux rope axis are generated (see Figs.~\ref{khnt4_3}b and Figure~\ref{khnt8_5}b).
These structures are lying mainly horizontally.

\section{Conclusions and Discussion}

The time evolution of a cold prominence initially embedded in a magnetic flux rope has
revealed interesting characteristics of the structure. We have concentrated on sustained
prominences, i.e., detached from the photosphere. The different numerical simulations have
shown how the magnetic field evolves in response to the dense prominence that is pulled
downwards by gravity.  The various twisted flux ropes considered in this work, together with
the embedded prominences, show a different evolution. We have shown that in some cases they
are clearly inclined towards a stationary state. These quasi-stationary three-dimensional
solutions can be used in the future for other purposes such as the analysis of winking
filaments produced by external perturbations like Moreton waves, EIT waves or nearby flares
and CMEs.

During the relaxation to the quasi-stationary equilibrium the system goes
through several periodic vertical oscillations. In the linear regime, i.e., when
the amplitude of oscillation is small in comparison with the local Alfv\'en
speed, the process of continuum damping transfers the excess of energy
in the system to the PCTR. It is in this highly inhomogeneous layer where the
process of phase-mixing takes place. This energy redistribution explains the
attenuation of the position of the center of mass of the prominence body.
Thus, at least in the linear regime the resonant absorption process is crucial
in the relaxation of the structure. The energy in the PCTR is
eventually dissipated by the numerical scheme used to perform the simulation,
and this dissipation can be hardly avoided due to the continuous generation of
small spatial scales with time by the phase-mixing mechanism. Fortunately,
although the numerical dissipation produces an energy loss, this does not
necessarily mean that the attenuation of the CM is dominated by this artificial
dissipation. In fact it is known that a small dissipation does not change the
damping time \citep{poedtskerner91} of the quasimode (i.e., the resonantly
damped global mode), it affects the dissipation in the layer but not the global
attenuation \citep[see][for the linear time evolution with
dissipation]{terretal06b}. The different convergence tests performed in the
present study point in this direction. Wave leakage is not relevant in our model
and does not contribute to the attenuation of the position of the CM.

An attractive alternative to the classical interpretation of the quasimode
is based on the proper superposition of Alfv\'en continuum modes
\citep[][]{cally1991,mannetal1995}. Recently, \citet{solerterr2015} have
investigated the generation of small scales in nonuniform solar magnetic flux
tubes. Using a modal expansion the initial global MHD transverse displacement
has been expressed as a superposition of Alfv\'en continuum modes that are
phase-mixed as time evolves. The comparison with the results of the quasimode
indicate that the modal analysis is more intelligible from the physical point of
view since it describes both the damping of global transverse motions and the
building up of small scales due to phase-mixing. It is not surprising that the
term phase-mixing has been also used in the past to refer to the conversion of
energy from large to small scales \citep[see][]{tataronisgross1973}.

The problem studied here demonstrates the necessity of a proper calculation of
the Alfv\'en and the slow continuum in general 3D geometries like the ones
considered in the present work.  The efforts done so far in the calculation of
the eigenproblem for magnetic field resonances in the context of magnetospheric
plasmas
\citep[see][]{singetal1981,rankinetal2006,kabinetal2007,degelingetal2010} may
shed some light about the application of the known techniques for closed
geomagnetic field lines to the problem of coronal loops and solar prominences.

The vertical oscillations found in the simulations are closely related to the
eigenmodes (quasimodes) of the configuration, and therefore, the same periodicities
should be found by perturbing (in the vertical direction) the stationary equilibrium.
Due to the symmetry in the system only the vertically polarized mode has been excited
in our numerical experiments, but according to previous works in similar toroidal
configurations without magnetic twist \citep[][]{terretal06a,vandetal09a} there should
be also a horizontally polarized mode with essentially the same period as the vertical
mode (this is true in the thin tube approximation only). Interestingly, the period of
the standing vertical oscillation does not change much with magnetic twist, at least
for the three configurations studied here. This can have implications from the
seismological point of view. The slow antisymmetric mode has been also detected in the
simulations but future studies are required to understand the existence of other modes
that due to the symmetry of the system have not been excited in our numerical
experiments.

Nonlinearity complicates the physics and thus the interpretation of the outcome
of the simulations. We have demonstrated that for the highest spatial resolution
($300\,\rm km$) the three twisted flux ropes analyzed in this work develop KH
instabilities at the PCTR. For all the cases, the attenuation of the center of
mass of the prominence is rather strong and the shear instability may contribute
to this attenuation (in the linear regime it is only due to resonant absorption). A
meticulous investigation of the effect of the KHI on the damping is required. 

For the values of twist considered here, constrained by observations (the
maximum number of turns of the magnetic field is at most two or three for stable
configurations), the poloidal component of the magnetic field, that provides an
additional stabilizing magnetic force, is unable to completely suppress the
shear instability at the PCTR. The evolution of the hosted prominence inside the
flux rope suggests the formation of elongated structures lying essentially
horizontal and generated by the KHI. These results are in contrast to the situation found
in the dipped magnetic arcade studied by \citet{terradasetal2015} with the long
axis of the prominence situated essentially perpendicular to the magnetic field.
In that configuration it was found that the MRT is very efficient \citep[see
also][]{hillieretal2012a,hillieretal2012b}. Fingers and plumes evolve at the
bottom of the interface between the prominence and corona  producing essentially
vertical structures.  Ideally, the different morphological observed properties
of prominences regarding the presence of either horizontal or vertical
structuring could be used to provide hints about the geometry of the
corresponding underlying magnetic structure. Horizontal structures are more
consistent with magnetic flux ropes, while vertical structuring along the
longitudinal axis of the prominence, is associated with arcade configurations
according to our study.

There are two basic reasons for the absence of the MRT instability in our flux rope
configuration. The first reason, and the most important,  is the orientation of the
prominence with respect to the magnetic field. In the arcade model the prominence is
permeated by a perpendicular magnetic field (when magnetic shear is small) while in our flux
rope model there is always an important magnetic component along the prominence body. This
means that for perturbations along the longitudinal axis, the flux rope prominence is
intrinsically much more stable with respect to the MRT instability since magnetic forces
produce a stabilizing effect that is missing in the unsheared arcade configuration. Second,
for perturbations perpendicular to the longitudinal axis of the prominence, the localized
twist provides a magnetic component that tends again to stabilize the structure. This is
similar to the effect of shear in the arcade configuration, but with the difference that the
stabilizing component changes much more rapidly with position in the flux rope model. This
strong variation results in a much more stable configuration \citep[see for
example][]{rudermanterradas2014}.

Finally, it seems clear that line-tying conditions at the base of the corona are
too restrictive. These conditions are crucial to obtain suspended prominences, and make the
effect of gas pressure along the magnetic field lines to  play a very important role in the
support, specially for the cases of low magnetic twist without magnetic dips. However, from
observations it is quite obvious that there is a continuous interchange of material with the
chromosphere, which is completely missing in our simulations due the line-tying conditions.
In addition, observations also show that blobs of material falling in the direction to the
chromosphere do not seem to stop at a given height as one would expect with these boundary
conditions in ideal MHD. The incorporation of more realistic conditions in our models
including the chromosphere and the pertinent non-ideal effects, such as radiative losses and
conduction,  is of capital relevance.

\acknowledgements \small J.T. and R. S. acknowledge support from MINECO and UIB through
a Ram\'on y Cajal grant. The authors acknowledge support from MINECO and FEDER funds
through project AYA2014-54485-P by the Spanish MICINN and FEDER Funds. The authors also
thank Dr.~Titov for providing a Fortran routine to calculate the flux rope magnetic
configuration of the Titov \& Demoulin model. The Open Source VisIt is supported by the
Department of Energy with funding from the Advanced Simulation and Computing Program
and the Scientific Discovery through Advanced Computing Program.


\begin{thebibliography}{71}
\expandafter\ifx\csname natexlab\endcsname\relax\def\natexlab#1{#1}\fi

\bibitem[{{Antolin} {et~al.}(2015){Antolin}, {Okamoto}, {De Pontieu},
  {Uitenbroek}, {Van Doorsselaere}, \& {Yokoyama}}]{antolinetal2015}
{Antolin}, P., {Okamoto}, T.~J., {De Pontieu}, B., {Uitenbroek}, H., {Van
  Doorsselaere}, T., \& {Yokoyama}, T. 2015, \apj, 809, 72

\bibitem[{{Arregui} \& {Ballester}(2010)}]{arrball10}
{Arregui}, I., \& {Ballester}, J.~L. 2010, Space Science Reviews

\bibitem[{{Arregui} {et~al.}(2011){Arregui}, {Soler}, {Ballester}, \&
  {Wright}}]{arreguietal11}
{Arregui}, I., {Soler}, R., {Ballester}, J.~L., \& {Wright}, A.~N. 2011, \aap,
  533, A60

\bibitem[{{Arregui} {et~al.}(2008){Arregui}, {Terradas}, {Oliver}, \&
  {Ballester}}]{arreguiterretal08}
{Arregui}, I., {Terradas}, J., {Oliver}, R., \& {Ballester}, J.~L. 2008, \apjl,
  682, L141

\bibitem[{{Berenger}(1994)}]{berenger1994}
{Berenger}, J.-P. 1994, Journal of Computational Physics, 114, 185

\bibitem[{{Blokland} \& {Keppens}(2011)}]{bloklandkeppens11}
{Blokland}, J.~W.~S., \& {Keppens}, R. 2011, \aap, 532, A94

\bibitem[{{Cally}(1991)}]{cally1991}
{Cally}, P.~S. 1991, Journal of Plasma Physics, 45, 453

\bibitem[{{Chen} \& {Hasegawa}(1974)}]{chenhasegawa1974}
{Chen}, L., \& {Hasegawa}, A. 1974, Physics of Fluids, 17, 1399

\bibitem[{Childs {et~al.}(2012)Childs, Brugger, Whitlock, Meredith, Ahern,
  Pugmire, Biagas, Miller, Harrison, Weber, Krishnan, Fogal, Sanderson, Garth,
  Bethel, Camp, R\"{u}bel, Durant, Favre, \& Navr\'{a}til}]{VisIt}
Childs, H., {et~al.} 2012, in {High Performance Visualization--Enabling
  Extreme-Scale Scientific Insight}, 357--372

\bibitem[{{De Groof} \& {Goossens}(2002)}]{degroofgoossens02}
{De Groof}, A., \& {Goossens}, M. 2002, \aap, 386, 691

\bibitem[{{De Groof} {et~al.}(2002){De Groof}, {Paes}, \&
  {Goossens}}]{degroofetal02}
{De Groof}, A., {Paes}, K., \& {Goossens}, M. 2002, \aap, 386, 681

\bibitem[{{Degeling} {et~al.}(2010){Degeling}, {Rankin}, {Kabin}, {Rae}, \&
  {Fenrich}}]{degelingetal2010}
{Degeling}, A.~W., {Rankin}, R., {Kabin}, K., {Rae}, I.~J., \& {Fenrich}, F.~R.
  2010, Journal of Geophysical Research (Space Physics), 115, 10212

\bibitem[{{Goossens} {et~al.}(2002){Goossens}, {Andries}, \&
  {Aschwanden}}]{goossetal02}
{Goossens}, M., {Andries}, J., \& {Aschwanden}, M.~J. 2002, \aap, 394, L39

\bibitem[{{Goossens} {et~al.}(1992){Goossens}, {Hollweg}, \&
  {Sakurai}}]{goossensetal92}
{Goossens}, M., {Hollweg}, J.~V., \& {Sakurai}, T. 1992, \solphys, 138, 233

\bibitem[{{Grossmann} \& {Tataronis}(1973)}]{grossmanntataronis1973}
{Grossmann}, W., \& {Tataronis}, J. 1973, Zeitschrift fur Physik, 261, 217

\bibitem[{{Heyvaerts} \& {Priest}(1983)}]{heypri83}
{Heyvaerts}, J., \& {Priest}, E.~R. 1983, \aap, 117, 220

\bibitem[{{Hillier} {et~al.}(2012{\natexlab{a}}){Hillier}, {Berger}, {Isobe},
  \& {Shibata}}]{hillieretal2012a}
{Hillier}, A., {Berger}, T., {Isobe}, H., \& {Shibata}, K. 2012{\natexlab{a}},
  \apj, 746, 120

\bibitem[{{Hillier} {et~al.}(2012{\natexlab{b}}){Hillier}, {Isobe}, {Shibata},
  \& {Berger}}]{hillieretal2012b}
{Hillier}, A., {Isobe}, H., {Shibata}, K., \& {Berger}, T. 2012{\natexlab{b}},
  \apj, 756, 110

\bibitem[{{Hillier} \& {van Ballegooijen}(2013)}]{hilliervan13}
{Hillier}, A., \& {van Ballegooijen}, A. 2013, \apj, 766, 126

\bibitem[{{Hollweg}(1987)}]{hollweg87}
{Hollweg}, J.~V. 1987, \apj, 312, 880

\bibitem[{{Hollweg} \& {Yang}(1988)}]{hollyang88}
{Hollweg}, J.~V., \& {Yang}, G. 1988, \jgr, 93, 5423

\bibitem[{{Hood} {et~al.}(2013){Hood}, {Ruderman}, {Pascoe}, {De Moortel},
  {Terradas}, \& {Wright}}]{hoodetal2013}
{Hood}, A.~W., {Ruderman}, M., {Pascoe}, D.~J., {De Moortel}, I., {Terradas},
  J., \& {Wright}, A.~N. 2013, \aap, 551, A39

\bibitem[{{Hu}(2001)}]{hu2001}
{Hu}, F.~Q. 2001, Journal of Computational Physics, 173, 455

\bibitem[{{Ionson}(1978)}]{ionson78}
{Ionson}, J.~A. 1978, \apj, 226, 650

\bibitem[{{Kabin} {et~al.}(2007){Kabin}, {Rankin}, {Waters}, {Marchand},
  {Donovan}, \& {Samson}}]{kabinetal2007}
{Kabin}, K., {Rankin}, R., {Waters}, C.~L., {Marchand}, R., {Donovan}, E.~F.,
  \& {Samson}, J.~C. 2007, \planss, 55, 820

\bibitem[{{Kuperus} \& {Raadu}(1974)}]{kuperusraadu1974}
{Kuperus}, M., \& {Raadu}, M.~A. 1974, \aap, 31, 189

\bibitem[{{Lee} \& {Roberts}(1986)}]{leerob86}
{Lee}, M.~A., \& {Roberts}, B. 1986, \apj, 301, 430

\bibitem[{{Li} \& {Zhang}(2012)}]{lizhang2012}
{Li}, T., \& {Zhang}, J. 2012, \apjl, 760, L10

\bibitem[{{Low} \& {Zhang}(2004)}]{lowzhang04}
{Low}, B.~C., \& {Zhang}, M. 2004, \apj, 609, 1098

\bibitem[{{Luna} {et~al.}(2014){Luna}, {Knizhnik}, {Muglach}, {Karpen}, {},
  {Kucera}, \& {Uritsky}}]{lunaetal2014}
{Luna}, M., {Knizhnik}, K., {Muglach}, K., {Karpen}, J., {}, H., {Kucera},
  T.~A., \& {Uritsky}, V. 2014, \apj, 785, 79

\bibitem[{{Mackay} {et~al.}(2010){Mackay}, {Karpen}, {Ballester}, {Schmieder},
  \& {Aulanier}}]{mackayetal10}
{Mackay}, D.~H., {Karpen}, J.~T., {Ballester}, J.~L., {Schmieder}, B., \&
  {Aulanier}, G. 2010, \ssr, 151, 333

\bibitem[{{Mann} {et~al.}(1995){Mann}, {Wright}, \& {Cally}}]{mannetal1995}
{Mann}, I.~R., {Wright}, A.~N., \& {Cally}, P.~S. 1995, \jgr, 100, 19441

\bibitem[{{Okamoto} {et~al.}(2015){Okamoto}, {Antolin}, {De Pontieu},
  {Uitenbroek}, {Van Doorsselaere}, \& {Yokoyama}}]{okamotoetal2015}
{Okamoto}, T.~J., {Antolin}, P., {De Pontieu}, B., {Uitenbroek}, H., {Van
  Doorsselaere}, T., \& {Yokoyama}, T. 2015, \apj, 809, 71

\bibitem[{{Parchevsky} \& {Kosovichev}(2007)}]{parchkoso2007}
{Parchevsky}, K.~V., \& {Kosovichev}, A.~G. 2007, \apj, 666, 547

\bibitem[{{Pascoe} {et~al.}(2010){Pascoe}, {Wright}, \& {De
  Moortel}}]{pascoeetal10}
{Pascoe}, D.~J., {Wright}, A.~N., \& {De Moortel}, I. 2010, \apj, 711, 990

\bibitem[{{Poedts} \& {Kerner}(1991)}]{poedtskerner91}
{Poedts}, S., \& {Kerner}, W. 1991, Physical Review Letters, 66, 2871

\bibitem[{{Poedts} {et~al.}(1992){Poedts}, {Kerner}, {Goedbloed}, {Keegan},
  {Huysmans}, \& {Schwarz}}]{poedtsetal1992}
{Poedts}, S., {Kerner}, W., {Goedbloed}, J.~P., {Keegan}, B., {Huysmans},
  G.~T.~A., \& {Schwarz}, E. 1992, Plasma Physics and Controlled Fusion, 34,
  1397

\bibitem[{{Priest} {et~al.}(1989){Priest}, {Hood}, \& {Anzer}}]{priestetal89}
{Priest}, E.~R., {Hood}, A.~W., \& {Anzer}, U. 1989, \apj, 344, 1010

\bibitem[{{Rae}(1982)}]{rae1982}
{Rae}, I.~C. 1982, Plasma Physics, 24, 133

\bibitem[{{Rankin} {et~al.}(2006){Rankin}, {Kabin}, \&
  {Marchand}}]{rankinetal2006}
{Rankin}, R., {Kabin}, K., \& {Marchand}, R. 2006, Advances in Space Research,
  38, 1720

\bibitem[{{Ruderman}(2007)}]{ruderman07}
{Ruderman}, M.~S. 2007, \solphys, 246, 119

\bibitem[{{Ruderman} {et~al.}(2010){Ruderman}, {Goossens}, \&
  {Andries}}]{rudermanetal2010}
{Ruderman}, M.~S., {Goossens}, M., \& {Andries}, J. 2010, Physics of Plasmas,
  17, 082108

\bibitem[{{Ruderman} \& {Roberts}(2002)}]{rudrob02}
{Ruderman}, M.~S., \& {Roberts}, B. 2002, \apj, 577, 475

\bibitem[{{Ruderman} \& {Terradas}(2013)}]{rudermanterradas2013}
{Ruderman}, M.~S., \& {Terradas}, J. 2013, \aap, 555, A27

\bibitem[{{Ruderman} \& {Terradas}(2015)}]{ruderterr2015}
---. 2015, \aap, 580, A57

\bibitem[{{Ruderman} {et~al.}(2014){Ruderman}, {Terradas}, \&
  {Ballester}}]{rudermanterradas2014}
{Ruderman}, M.~S., {Terradas}, J., \& {Ballester}, J.~L. 2014, \apj, 785, 110

\bibitem[{{Ryutova} {et~al.}(2010){Ryutova}, {Berger}, {Frank}, {Tarbell}, \&
  {Title}}]{ryuetal10}
{Ryutova}, M., {Berger}, T., {Frank}, Z., {Tarbell}, T., \& {Title}, A. 2010,
  \solphys, 267, 75

\bibitem[{{Sakurai} {et~al.}(1991){Sakurai}, {Goossens}, \&
  {Hollweg}}]{sakuraietal91}
{Sakurai}, T., {Goossens}, M., \& {Hollweg}, J.~V. 1991, \solphys, 133, 227

\bibitem[{{Sedl{\'a}{\v c}ek}(1995)}]{sedla1995}
{Sedl{\'a}{\v c}ek}, Z. 1995, in American Institute of Physics Conference
  Series, Vol. 345, American Institute of Physics Conference Series, 119--126

\bibitem[{{Singer} {et~al.}(1981){Singer}, {Southwood}, {Walker}, \&
  {Kivelson}}]{singetal1981}
{Singer}, H.~J., {Southwood}, D.~J., {Walker}, R.~J., \& {Kivelson}, M.~G.
  1981, \jgr, 86, 4589

\bibitem[{{Soler} {et~al.}(2010{\natexlab{a}}){Soler}, {Arregui}, {Oliver}, \&
  {Ballester}}]{soleretal10}
{Soler}, R., {Arregui}, I., {Oliver}, R., \& {Ballester}, J.~L.
  2010{\natexlab{a}}, \apj, 722, 1778

\bibitem[{{Soler} {et~al.}(2009){Soler}, {Oliver}, {Ballester}, \&
  {Goossens}}]{soleretal09c}
{Soler}, R., {Oliver}, R., {Ballester}, J.~L., \& {Goossens}, M. 2009, \apjl,
  695, L166

\bibitem[{{Soler} \& {Terradas}(2015)}]{solerterr2015}
{Soler}, R., \& {Terradas}, J. 2015, \apj, 803, 43

\bibitem[{{Soler} {et~al.}(2010{\natexlab{b}}){Soler}, {Terradas}, {Oliver},
  {Ballester}, \& {Goossens}}]{soleretal2010}
{Soler}, R., {Terradas}, J., {Oliver}, R., {Ballester}, J.~L., \& {Goossens},
  M. 2010{\natexlab{b}}, \apj, 712, 875

\bibitem[{{Southwood}(1974)}]{southwood1974}
{Southwood}, D.~J. 1974, \planss, 22, 483

\bibitem[{{Tataronis} \& {Grossmann}(1973)}]{tataronisgross1973}
{Tataronis}, J., \& {Grossmann}, W. 1973, Zeitschrift fur Physik, 261, 203

\bibitem[{{Terradas} {et~al.}(2008{\natexlab{a}}){Terradas}, {Andries},
  {Goossens}, {Arregui}, {Oliver}, \& {Ballester}}]{terradasetal08}
{Terradas}, J., {Andries}, J., {Goossens}, M., {Arregui}, I., {Oliver}, R., \&
  {Ballester}, J.~L. 2008{\natexlab{a}}, \apjl, 687, L115

\bibitem[{{Terradas} {et~al.}(2008{\natexlab{b}}){Terradas}, {Arregui},
  {Oliver}, {Ballester}, {Andries}, \& {Goossens}}]{terrarretal08}
{Terradas}, J., {Arregui}, I., {Oliver}, R., {Ballester}, J.~L., {Andries}, J.,
  \& {Goossens}, M. 2008{\natexlab{b}}, \apj, 679, 1611

\bibitem[{{Terradas} \& {Goossens}(2012)}]{terradasgoossens2012}
{Terradas}, J., \& {Goossens}, M. 2012, \aap, 548, A112

\bibitem[{{Terradas} {et~al.}(2006{\natexlab{a}}){Terradas}, {Oliver}, \&
  {Ballester}}]{terretal06b}
{Terradas}, J., {Oliver}, R., \& {Ballester}, J.~L. 2006{\natexlab{a}}, \apj,
  642, 533

\bibitem[{{Terradas} {et~al.}(2006{\natexlab{b}}){Terradas}, {Oliver}, \&
  {Ballester}}]{terretal06a}
---. 2006{\natexlab{b}}, \apjl, 650, L91

\bibitem[{{Terradas} {et~al.}(2015){Terradas}, {Soler}, {Luna}, {Oliver}, \&
  {Ballester}}]{terradasetal2015}
{Terradas}, J., {Soler}, R., {Luna}, M., {Oliver}, R., \& {Ballester}, J.~L.
  2015, \apj, 799, 94

\bibitem[{{Titov} \& {D{\'e}moulin}(1999)}]{titovdem1999}
{Titov}, V.~S., \& {D{\'e}moulin}, P. 1999, \aap, 351, 707

\bibitem[{{T{\"o}r{\"o}k} \& {Kliem}(2003)}]{torokkliem2003}
{T{\"o}r{\"o}k}, T., \& {Kliem}, B. 2003, \aap, 406, 1043

\bibitem[{{T{\"o}r{\"o}k} {et~al.}(2004){T{\"o}r{\"o}k}, {Kliem}, \&
  {Titov}}]{toroketal2004}
{T{\"o}r{\"o}k}, T., {Kliem}, B., \& {Titov}, V.~S. 2004, \aap, 413, L27

\bibitem[{{Tripathi} {et~al.}(2009){Tripathi}, {Isobe}, \&
  {Jain}}]{tripatetal09}
{Tripathi}, D., {Isobe}, H., \& {Jain}, R. 2009, \ssr, 149, 283

\bibitem[{{van Doorsselaere} {et~al.}(2009){van Doorsselaere}, {Verwichte}, \&
  {Terradas}}]{vandetal09a}
{van Doorsselaere}, T., {Verwichte}, E., \& {Terradas}, J. 2009, \ssr, 149, 299

\bibitem[{{Wright}(1992)}]{wrigth1992}
{Wright}, A.~N. 1992, \jgr, 97, 6439

\bibitem[{{Wright} \& {Rickard}(1995)}]{wrightrick1995}
{Wright}, A.~N., \& {Rickard}, G.~J. 1995, \apj, 444, 458

\bibitem[{{Xia} {et~al.}(2014{\natexlab{a}}){Xia}, {Keppens}, {Antolin}, \&
  {Porth}}]{xiaetal2014b}
{Xia}, C., {Keppens}, R., {Antolin}, P., \& {Porth}, O. 2014{\natexlab{a}},
  \apjl, 792, L38

\bibitem[{{Xia} {et~al.}(2014{\natexlab{b}}){Xia}, {Keppens}, \&
  {Guo}}]{xiaetal2014a}
{Xia}, C., {Keppens}, R., \& {Guo}, Y. 2014{\natexlab{b}}, \apj, 780, 130

\end{thebibliography}
\end{document}